\documentclass{article}

\usepackage{arxiv}

\usepackage[utf8]{inputenc}
\usepackage[T1]{fontenc}
\usepackage{url}
\usepackage{booktabs}
\usepackage{amsmath,amssymb,amsfonts}
\usepackage{nicefrac}
\usepackage{microtype}
\usepackage{graphicx}
\usepackage[numbers,square,sort&compress]{natbib}
\usepackage{xspace}
\usepackage{xcolor}
\usepackage{colortbl}
\usepackage{enumitem}
\usepackage[singlelinecheck=false]{caption}
\usepackage{subcaption}
\usepackage{tabularx}
\usepackage{multirow}
\usepackage{multicol}
\usepackage{threeparttable}
\usepackage{makecell}
\usepackage{longtable}
\usepackage{float}
\usepackage[bottom]{footmisc}
\usepackage{hyperref}
\usepackage{doi}
\usepackage{cleveref}

\usepackage{needspace}
\usepackage{etoolbox}
\pretocmd{\section}{\needspace{8\baselineskip}}{}{}
\pretocmd{\subsection}{\needspace{6\baselineskip}}{}{}
\pretocmd{\subsubsection}{\needspace{5\baselineskip}}{}{}
\pretocmd{\paragraph}{\needspace{4\baselineskip}}{}{}

\newcommand{\Description}[1]{}

\DeclareRobustCommand{\DEL}[1]{}

\newlist{hyp}{enumerate}{2}
\setlist[hyp,1]{before=\itshape,font=\itshape, label=Hypothesis \arabic*:,ref= \arabic*, leftmargin=*}
\setlist[hyp,2]{before=\itshape,font=\itshape, label=Hypothesis \arabic*:,ref= \arabic*, leftmargin=*}
\newlist{req}{enumerate}{2}
\setlist[req,1]{label=RQ \arabic*:,ref= \textbf{\arabic*}, leftmargin=*}

\title{Think Once, Reflect Twice: How Feedback Shapes Human-AI Collaboration}

\author{
  \href{https://orcid.org/0009-0005-3885-8365}{\includegraphics[scale=0.06]{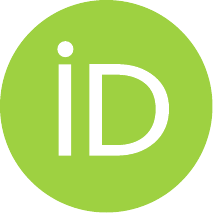}\hspace{1mm}Joshua Holstein}$^{*}$ \\
  Karlsruhe Institute of Technology\\
  Karlsruhe, Germany \\
  \texttt{joshua.holstein@kit.edu} \\
  \And
  \href{https://orcid.org/0000-0002-9378-0872}{\includegraphics[scale=0.06]{orcid.pdf}\hspace{1mm}Philipp Spitzer}$^{*}$ \\
  Karlsruhe Institute of Technology\\
  Karlsruhe, Germany \\
  \texttt{philipp.spitzer@kit.edu} \\
  \And
  \href{https://orcid.org/0009-0009-7782-7528}{\includegraphics[scale=0.06]{orcid.pdf}\hspace{1mm}Sebastian Sch\"afer} \\
  Karlsruhe Institute of Technology\\
  Karlsruhe, Germany \\
  \texttt{sebastian.schaefer2@kit.edu} \\
  \AND
  \href{https://orcid.org/0009-0000-5739-9241}{\includegraphics[scale=0.06]{orcid.pdf}\hspace{1mm}Maximilian F\"orster} \\
  Karlsruhe Institute of Technology\\
  Karlsruhe, Germany \\
  \texttt{maximilian.foerster@kit.edu} \\
  \And
  \href{https://orcid.org/0000-0001-8731-654X}{\includegraphics[scale=0.06]{orcid.pdf}\hspace{1mm}Gerhard Satzger} \\
  Karlsruhe Institute of Technology\\
  Karlsruhe, Germany \\
  \texttt{gerhard.satzger@kit.edu} \\
}

\renewcommand{\shorttitle}{Think Once, Reflect Twice}

\hypersetup{
colorlinks=true, linkcolor=blue!60!black, citecolor=blue!60!black, urlcolor=blue!60!black,
pdftitle={Think Once, Reflect Twice: How Feedback Shapes Human-AI Collaboration},
pdfsubject={cs.HC},
pdfauthor={Joshua Holstein, Philipp Spitzer, Sebastian Sch\"afer, Maximilian F\"orster, Gerhard Satzger},
pdfkeywords={Human-AI Collaboration, Reflection, Appropriate Reliance, Artificial Intelligence, Feedback},
}

\begin{document}
\maketitle
{\renewcommand{\thefootnote}{}\footnotetext{$^{*}$Both authors contributed equally to this research.}}

\begin{abstract}
As artificial intelligence (AI) increasingly supports human decision-making in high-stakes domains, designing effective human-AI collaboration becomes critical. Successful collaboration requires humans to reflect on decision-making processes and adjust their strategies accordingly. Current research typically investigates interventions, such as explainable AI, that target reflection during the decision-making process (reflection-in-action). Less attention has been given to retrospective examination of decisions (reflection-on-action). This work explores reflection-on-action by studying the impact of feedback on reliance behavior and human-AI team performance. Specifically, we investigate feedback interventions targeting overrides of AI advice through a behavioral study (N=244). We find that such feedback influences reflection-on-action, while unidirectional feedback risks reducing subsequent reflection-in-action. Furthermore, we identify opposing shifts in reliance behavior that leave overall human-AI team performance unchanged but show context-dependent effects. Balanced feedback did not show these risks, preserving both concurrent and retrospective reflection, offering insights for designing for appropriate reliance on AI.
\end{abstract}


\section{Introduction}
\label{sec-introduction}
\textit{Envision you are collaborating with an artificial intelligence (AI) as a loan officer to evaluate the creditworthiness of loan applicants. When the AI advised rejection but you favored approval (or vice versa), who would you rely on? How would your reliance on the AI shift after receiving feedback on the outcomes of decisions you overrode?}

AI systems increasingly support human decision-making across domains \citep{herath2022adoption, rahman2023adoption}, including high-stakes applications such as medical diagnosis \citep{varghese2020artificial}, legal advice \citep{zahra2025regulating}, and loan approval \citep{purificato2023use}. In these applications, humans no longer decide alone but collaborate with an AI system that advises them. Besides understanding the task and its domain, they therefore also need to understand the AI advice and develop strategies for when to follow it. When this collaboration fails, the people affected by the decisions bear the consequences, for example, for their health in medical diagnosis \citep{aldosari2025challenges}, their liberty in recidivism decisions \citep{angwin2016machine}, and their access to credit in loan decisions \citep{saxena2019fairness}.

These high-stakes consequences underscore the need for humans to develop collaboration strategies with AI systems that enable appropriate reliance, i.e., the ability to effectively complement one's own judgment with available AI advice \citep{schemmer2023appropriate, schoeffer2024explanations}. If humans blindly rely on AI advice, they risk systematic overreliance \citep{buccinca2021trust, schemmer2022influence} and deskilling \citep{morrison2024impact, natali2025ai}; if they ignore AI advice, they miss out on potential improvements \citep{schoeffer2025ai}. Developing effective collaboration strategies to achieve appropriate reliance requires \textit{reflection}. In general terms, reflection is the purposeful and critical analysis of one's own experiences, knowledge, and actions, aimed at a deeper understanding that informs future action \citep{boud1985reflection, schon1987educating}. In human-AI collaboration, this analysis extends to the advice of the AI: it helps humans recognize systematic errors, their own and the AI's, and learn from them \citep{danry2025human}. Research on reflection distinguishes between two equally important types of reflection that occur at two different points in time: \textit{reflection-in-action} that occurs during decision-making, as deliberate reasoning about the case and the advice at hand, and \textit{reflection-on-action} that occurs retrospectively, targeting past decision-making and the lessons drawn from its outcomes \citep{schon1987educating}. Research in human-AI collaboration has so far mainly focused on reflection-in-action: for instance, researchers have investigated whether explanations \citep{li2025text, solano2024explorable} or uncertainty displays \citep{zhang2020effect, holstein25balancing} along with AI advice can support humans in critically assessing AI advice. While reflection-in-action can support understanding of specific advice \citep{schoeffer2024explanations, buccinca2025contrastive} and may even enable implicit learning of effective collaboration strategies over time \citep{ma2024you, zahn2025calibration}, reflection-on-action is equally important to calibrate collaboration strategies. 

Reflection-on-action can build on what reflection-in-action lacks, the outcomes of past decisions. In real-world decision-making, whether a decision was correct is unknown when it is made \citep{gajos2022learning} and becomes apparent only later, once its outcome can be observed, often after a substantial delay. For example, a loan officer might approve a loan though the AI advised rejection, thereby overriding the AI advice. Later, the loan officer would likely learn whether the loan was repaid and could then retrospectively reflect on whether and why their own judgment or the AI advice was correct. Such \textit{feedback on past overrides} provides information on whether (and why) the user’s overrides of AI advice were successful (i.e., led to better outcomes) or unsuccessful (i.e., led to worse outcomes). Feedback on past overrides can be applied to any situation in which, after a decision is made, its actual outcome can be observed. Feedback in general has been attributed an important role in HCI literature \citep{otero2022cognitive}. For instance, \citet{foerster2025feedback} and  \citet{spitzer2024transferring} investigated the effectiveness of feedback provided through AI on human learning. \citet{ma2024you} explored the impact of calibration-status feedback, i.e., feedback indicating whether the accuracy of human decisions matches their confidence, on reliance. However, the role of feedback on past overrides, i.e., on whether disagreeing with the AI advice led to superior outcomes, has so far been overlooked. We aim to address this important gap by posing the following research questions (RQ):

\begin{req}[leftmargin=1.06cm, labelindent=0pt, labelwidth=0em, label=\textbf{RQ\arabic*}:, ref=\arabic*]
    \item How does feedback on the outcomes of past overrides of AI advice affect human reflection? \label{rq1}
    \item How does feedback on the outcomes of past overrides of AI advice affect subsequent reliance behavior and human-AI team performance? \label{rq2}
\end{req}

To address these questions, we conducted a behavioral experiment in which participants evaluated loan applications with the support of an AI system over three rounds. Drawing on theory on reflection \citep{schon1987educating}, reliance behavior \citep{schoeffer2024explanations}, and feedback effectiveness \citep{hattie2007power, narciss2008feedback}, we designed interventions that provided participants with systematic feedback on their (un)successful overrides of AI advice after the first two rounds. Specifically, participants received feedback on past overrides of AI advice that led to (i) correct decisions (successful), (ii) incorrect decisions (unsuccessful), or (iii) both correct and incorrect decisions (balanced), or (iv) they received no feedback in our control condition.

Our findings indicate that feedback on past overrides can induce reflection-on-action. However, for our unidirectional feedback conditions (only successful or unsuccessful overrides), we find lower subsequent reflection-in-action, significantly so after unsuccessful overrides. These cognitive changes accompany changes in reliance behavior, where feedback on successful overrides is associated with marginally lower adherence to AI advice, while feedback on unsuccessful overrides is associated with more adherence to incorrect AI advice. Although these behavioral changes do not translate into significant changes in overall human-AI team performance, subgroup analysis reveals context-dependent differences. Our findings make three important contributions: (1) We address a critical gap in the literature by introducing interventions that allow the investigation of reflection-on-action in human-AI collaboration, complementing the growing body of work that predominantly focuses on reflection-in-action. (2) Drawing on reliance theory, we translate established frameworks into feedback interventions on past overrides that provide systematic information about past collaboration outcomes and that move explanations from the moment of decision to the review of past overrides. (3) We show that feedback on past overrides can increase reflection-on-action and is associated with changes in reliance behavior, indicating that the two forms of reflection interact and that selective feedback can itself miscalibrate reliance, insights that inform future human-AI collaboration system design.
\section{Foundations and Related Work}
\label{sec:background}
\textbf{Reliance and Reflection in Human-AI Collaboration.}
Research on human-AI collaboration has extensively investigated the reliance behavior of humans in AI-assisted decision-making scenarios \citep{schemmer2023appropriate, schoeffer2024explanations, schoeffer2025ai}. Ideally, humans selectively follow AI advice when it is correct and override it when it is incorrect. However, a substantial body of work has documented the persistent problem of overreliance, where humans systematically follow AI advice even when their independent judgment would yield superior outcomes \citep{vaccaro2024combinations, hemmer2021human}. Conversely, underreliance occurs when humans systematically dismiss and override AI advice, even in situations where such advice would improve decision quality. \citet{ma2024you} show that such inappropriate reliance often emerges from a mismatch between humans’ self-confidence and their perception of the AI’s competence. Thus, calibration of humans’ perception of their own abilities and the AI's abilities is essential for appropriate reliance \citep{ma2024you}.

A key mechanism towards appropriate reliance is reflective practice in decision-making \citep{buccinca2021trust, ma2025towards}.  An important distinction is the timing of reflection within the decision-making process. \citet{schon1987educating} distinguishes between \textit{reflection-in-action} and \textit{reflection-on-action}. Reflection-in-action is a situational, real-time form of reflection that occurs during the decision-making process and before a decision has been made, enabling immediate reframing and adjustment of cognitive approaches. In contrast, reflection-on-action refers to the retrospective examination of decision-making, in which individuals analyze their actions after completing a task, thereby enabling the reframing and adjustment of cognitive approaches for future decisions. 

Reflective practice in human-AI collaboration has gained growing attention in literature \citep{abdelkarim2023reflection, bentvelzen2022revisiting}. For instance, \citet{taudien2024metacognition} highlight the critical role of metacognition, describing humans’ ability to reflect on their own cognitive approaches, including perception and decision-making. Ideally, reflection leads to calibration of humans’ perception of their own abilities and the AI's abilities and, accordingly, adjustment of reliance behavior \citep{zahn2025calibration}. However, even if humans exhibit a high degree of metacognition, they can still be prone to cognitive shortcuts, which can lead to poor decisions \citep{he2023knowing}. Thus, HCI research has explored how interactive systems, such as those used in human-AI collaboration, can prompt reflection, demonstrating the need for deliberate ``breakdowns'' (i.e., situations of surprise) and support of ``inquiry'' to trigger reflection \citep{cho2022reflection, fleck2010reflecting, baumer2015reflective}. \citet{abdelkarim2023reflection} have brought forward the idea that AI can serve as a reflection partner, proposing a grounded, theory-informed model of AI-induced reflection.

\textbf{Reflection Mechanisms in Human-AI Collaboration.}
So far, research has examined initial mechanisms that trigger reflection in human-AI collaboration, as shown in Table~\ref{tab:xai_interventions}. Researchers have predominantly used experimental studies to investigate mechanisms such as explanations and confidence displays alongside AI advice.

Findings suggest that the provision of confidence or uncertainty displays can improve trust calibration and have positive effects on humans' reliance behavior \citep{zhang2020effect, walter2024follow}. Researchers have also identified positive effects of explanations on trust and confidence calibration \citep{balakrishnan2025human}, especially when they were applied in combination with other mechanisms, such as cognitive forcing interventions and case-specific confidence information \citep{buccinca2021trust, gentile2025counterfactual, zhang2020effect, walter2024follow}. Critically, the framing and presentation of uncertainty \citep{holstein25balancing} and explanations \citep{danry2023tell, gentile2025counterfactual} can influence their effectiveness. Existing literature also indicated that the impact of reflection mechanisms is highly dependent on contextual and individual factors. \citet{Bansal2019} highlight mental models as a key human factor and investigate the effect of key properties of AI and tasks on mental models. In the study by \citet{spitzer2025imperfections}, humans’ cognitive styles and verbal ability moderated how effectively they processed explanations and reflected on AI advice. Accordingly, researchers emphasize the importance of communication style \citep{ma2025towards, danry2023tell}. For instance, deliberative AI---enabling humans to discuss with the AI---was found more effective for appropriate reliance than explainable AI \citep{ma2025towards}. Similarly, \citet{danry2023tell} found that explanations that are framed as questions have a positive effect. While updates on the side of the AI model in human-AI teams, as well, can trigger reflection, \citet{bansal2019updates} claim the need for compatibility of updates with mental models. \citet{ma2024you} specifically studied mechanisms for confidence calibration using different interventions, such as asking for opposite arguments, thinking in bets, and calibration status feedback. They find that all three interventions have a positive influence on human performance and reduce under-reliance. 

In sum, the literature emphasizes the critical role of reflection mechanisms, primarily the application of explainable AI and confidence displays, in calibrating humans' perception of their own abilities and AI's abilities and adjusting their reliance behavior. Most existing studies examine mechanisms that take effect alongside AI advice before a decision is made, which \citet{schon1987educating} call reflection-in-action. The few studies that also address reflection-on-action examine calibration-status feedback on humans' confidence \citep{ma2024you} or updates of the AI model \citep{bansal2019updates}, and neither gives humans feedback on the outcomes of their own overrides of AI advice. So far, the role of reflection-on-action, i.e., the retrospective evaluation of decision-making in collaboration with the AI, remains underexplored in human-AI collaboration research. In particular, feedback on past overrides as a potential reflection mechanism influencing reliance behavior has so far been neglected. We aim to address this gap.

\begin{longtable}{p{2.8cm} p{3.6cm} p{3.2cm} p{4.2cm}}
\caption{Literature investigating the role of reflection mechanisms for appropriate reliance in human-AI collaboration}
\label{tab:xai_interventions}\\
\toprule
\textbf{Paper} & \textbf{Methodology} & \textbf{Reflection mechanism} & \textbf{Findings} \\
\midrule
\endfirsthead
\caption[]{Literature investigating the role of reflection mechanisms for appropriate reliance in human-AI collaboration (continued)}\\
\toprule
\textbf{Paper} & \textbf{Methodology} & \textbf{Reflection mechanism} & \textbf{Findings} \\
\midrule
\endhead

\midrule
\multicolumn{4}{r}{\small Continued on next page}
\\\bottomrule
\endfoot

\bottomrule
\endlastfoot

\citet{zhang2020effect} &
Quantitative study on confidence displays and local explanations on trust-calibration &
Confidence displays \& explanations \newline
\textit{Reflection-in-action} &
Case-specific confidence information combined with explanations improved trust calibration compared to explanations alone. \\

\citet{ma2024you} &
Quantitative study investigating the effects of human self-confidence calibration  &
Interventions: think opposite, thinking in bets, calibration status feedback \newline
\textit{Reflection-in-action \newline\& reflection-on-action} &
Confidence calibration mechanisms reduced under-reliance and have a positive influence on human performance \\

\citet{holstein25balancing} &
Quantitative study investigating the effects of uncertainty framing and decomposition on human reliance in AI systems &
Uncertainty displays \newline
\textit{Reflection-in-action} &
Framing and presentation of uncertainty influence users’ reliance behavior \\

\citet{gentile2025counterfactual} &
Quantitative study examining the effects of counterfactual explanations combined with other explanation types  &
Counterfactual explanations \newline
\textit{Reflection-in-action} &
Adding Counterfactual explanations to other explanation types reduced false negatives while reducing decision time and workload. \\

\citet{danry2023tell} &
Quantitative study on the effects of explanations framed as questions &
Questioning explanations \newline
\textit{Reflection-in-action} &
AI-framed questioning explanations increased human discernment over causal explanations \\

\citet{spitzer2025imperfections} & 
Quantitative study investigating individual differences in explanation processing &
Imperfect explanations \newline
\textit{Reflection-in-action} &
Cognitive style and verbal ability moderated explanation effectiveness; imperfect explanations had varying effects depending on human characteristics. \\

\citet{walter2024follow} &
Quantitative study investigating the impact of confidence scores and explanations on appropriate reliance &
Confidence displays \& explanation \newline
\textit{Reflection-in-action} &
Confidence displays alongside explanations reduced over-reliance but increased under-reliance \\

\citet{buccinca2021trust} & 
Quantitative study investigating cognitive forcing interventions to compel people to engage more thoughtfully with explanations &
Cognitive forcing functions \newline
\textit{Reflection-in-action} &
Cognitive forcing functions alongside explanations reduced overreliance. \\

\citet{ma2025towards} &
Mixed-methods study designing and evaluating deliberative AI &
Deliberative AI \newline
\textit{Reflection-in-action} &
Deliberative AI was more effective than explainable AI for appropriate reliance \\

\citet{balakrishnan2025human} &
Quantitative study investigating measures against naive advice weighting in favor of users using their private information &
Feature transparency explanations \& nudging \newline
\textit{Reflection-in-action} &
Feature transparency explanations and nudging reduced naive advice weighting \\

\citet{bansal2019updates} &
Quantitative study investigating how updates (increasing AI performance) affect team performance &
Model updates \newline
\textit{Reflection-in-action \newline\& reflection-on-action} &
Updates that are incompatible with mental models can hurt team performance \\

\end{longtable}

\section{Theoretical Development}
\label{sec:theoretical_development}

Reflection has been established as a cornerstone for effective learning and performance improvement across domains~\citep{schon1987educating}. In education, reflection enhances learning outcomes by helping students identify knowledge gaps and adjust learning strategies~\citep{hattie2007power}. In professional contexts, reflection enables practitioners to recognize patterns in their decision-making and calibrate their expertise over time~\citep{vince2009reflection}. The effectiveness of reflection stems from its ability to activate metacognitive processes that help individuals recognize and correct systematic biases in their reasoning~\citep{flavell1979metacognition}.

These benefits of reflection extend to human-AI collaboration, where users must critically evaluate both their own judgment and AI advice to make effective decisions. Prior work supports this evaluation during decision-making, for instance through explanations \citep{Lundberg2017, solano2024explorable} and cognitive forcing functions \citep{buccinca2021trust}. We instead target reflection-on-action, which becomes possible once the ground truth of past decisions is available. We assume that feedback on past overrides is a critical component for effective reflection-on-action, as it provides information on whether and why past decision-making was successful \citep{hattie2007power, narciss2008feedback}. 

In human-AI collaboration, cognitive approaches mainly refer to humans' collaboration strategies, i.e., strategies humans develop about whether they should adhere to AI advice or override it. Thus, reflection-on-action translates into reflection on when and why humans successfully or unsuccessfully override AI advice, with the aim of learning from errors for future collaboration \citep{tulis2016learning}. We assume that providing feedback on past overrides, which reveals when and why AI advice was correct or incorrect compared to the human's decision, facilitates such cognitive engagement. Therefore, we formulate our first hypothesis:

\begin{hyp}[wide, leftmargin=0cm, labelindent=0pt, labelwidth=0em]
\item Feedback on past overrides, revealing when and why the AI diverged from the human’s decision, will increase reflection-on-action.
\label{hyp1}
\end{hyp}

Organizational~\citep{prue1981performance} and psychological research~\citep{kluger1996effects} suggest that feedback can improve human performance when communicated appropriately. Research demonstrates that feedback enhances individuals' awareness of decision consequences in complex settings, enabling them to identify and adopt more efficient alternatives \citep{casal2017feedback}. In collaborative decision-making contexts, such as human-AI collaboration, this mechanism translates into how individuals calibrate their reliance on different information sources, such as AI advice, based on the outcomes of past decision-making \citep{bailey2023meta}. When individuals receive information about the consequences of adhering to or overriding advice, they update their beliefs about when to rely on their own judgment versus external input. We thus expect that feedback interventions also translate into behavioral changes in collaborative decision-making settings \citep{pescetelli2021role}.  We propose that feedback on past overrides can influence reliance calibration by providing information about when and why AI advice was correct or incorrect compared to the human's judgment. Specifically, feedback on unsuccessful overrides should increase reliance by revealing missed opportunities, while feedback on successful overrides should decrease reliance by reinforcing appropriate skepticism, leading to our second hypothesis:

\begin{hyp}[resume, wide, leftmargin=0cm, labelindent=0pt, labelwidth=0em]
\item Humans who receive feedback on their unsuccessful past overrides of AI advice will show increased reliance on AI advice, while humans who receive feedback on their successful past overrides of AI advice will show lower reliance on AI advice.
\label{hyp2}
\end{hyp}

Appropriate reliance behavior directly translates into improved human-AI team performance, as better calibrated collaboration strategies enable such teams to leverage their complementary strengths~\citep{schemmer2023appropriate, hemmer2025complementarity}. If feedback on past overrides successfully improves reliance calibration, we should observe corresponding performance improvements.
Meta-analyses provide evidence that feedback interventions improve performance when they focus on task-relevant processes rather than self-evaluation \citep{kluger1996effects}.  Educational research further confirms that feedback highlighting both successful and unsuccessful decision patterns enables learners to develop more accurate mental models of task requirements \citep{shute2008focus}. We assume that, in human-AI collaboration, feedback on unsuccessful overrides will decrease missed opportunities, while feedback on successful overrides will reduce the adoption of mistakes from AI advice in future decisions. Combining both assumptions, feedback on past overrides that contains both humans' successful and unsuccessful overrides (i.e., balanced feedback) should result in even higher performance gains. Accordingly, we hypothesize:

\begin{hyp}[resume, wide, leftmargin=0cm, labelindent=0pt, labelwidth=0em]
\item Humans who receive feedback on their past overrides of the AI advice (feedback on successful overrides, feedback on unsuccessful overrides, and balanced feedback) achieve higher human-AI team performance, with balanced feedback leading to the highest performance gains.
\label{hyp3}
\end{hyp}

\section{Experimental Design}
\label{sec:experimental_design}

Our aim is to provide initial insights into the underexplored area of how feedback on past overrides of AI advice affects human-AI collaboration. To this end, we conduct an IRB-approved behavioral experiment examining both subjective and objective variables on reflection, reliance behavior, and human-AI team performance. For our experiment, we instantiated an AI system and an XAI method to generate AI advice and explanations based on real data on loan applications. We systematically manipulate the feedback on past overrides, resulting in four experimental groups: two groups receiving unidirectional feedback (either successful overrides or unsuccessful overrides), one group receiving balanced feedback (successful and unsuccessful overrides), and a control group receiving no feedback. The study follows a convergent mixed-methods design with quantitative priority \citep{fetters2013achieving}. The experiment tests our hypotheses, and the open-ended responses, collected in the same post-task questionnaire, serve to explain and extend the quantitative results. 

\textbf{Task and Dataset.} Following established practice in HCI research on human-AI collaboration, we selected loan approval as our experimental task \citep{yurrita2023disentangling, kapania2022because, van2021effect}, specifically the Home Credit Default Risk dataset from Kaggle \citep{kaggleHomeLoan}. The task involves evaluating multiple complex factors, such as income, credit history, and debt-to-income ratios, that benefit from algorithmic support while remaining accessible to participants even without extensive domain training. Based on feature importance rankings from a decision tree trained on the full dataset, we select eight key features: demographic characteristics (gender, marital status, education, number of dependents), financial indicators (household income, loan amount, loan duration), and credit history.

\begin{figure}[htbp]
    \centering
    \begin{subfigure}[b]{0.48\textwidth}
        \centering
        \includegraphics[width=\linewidth]{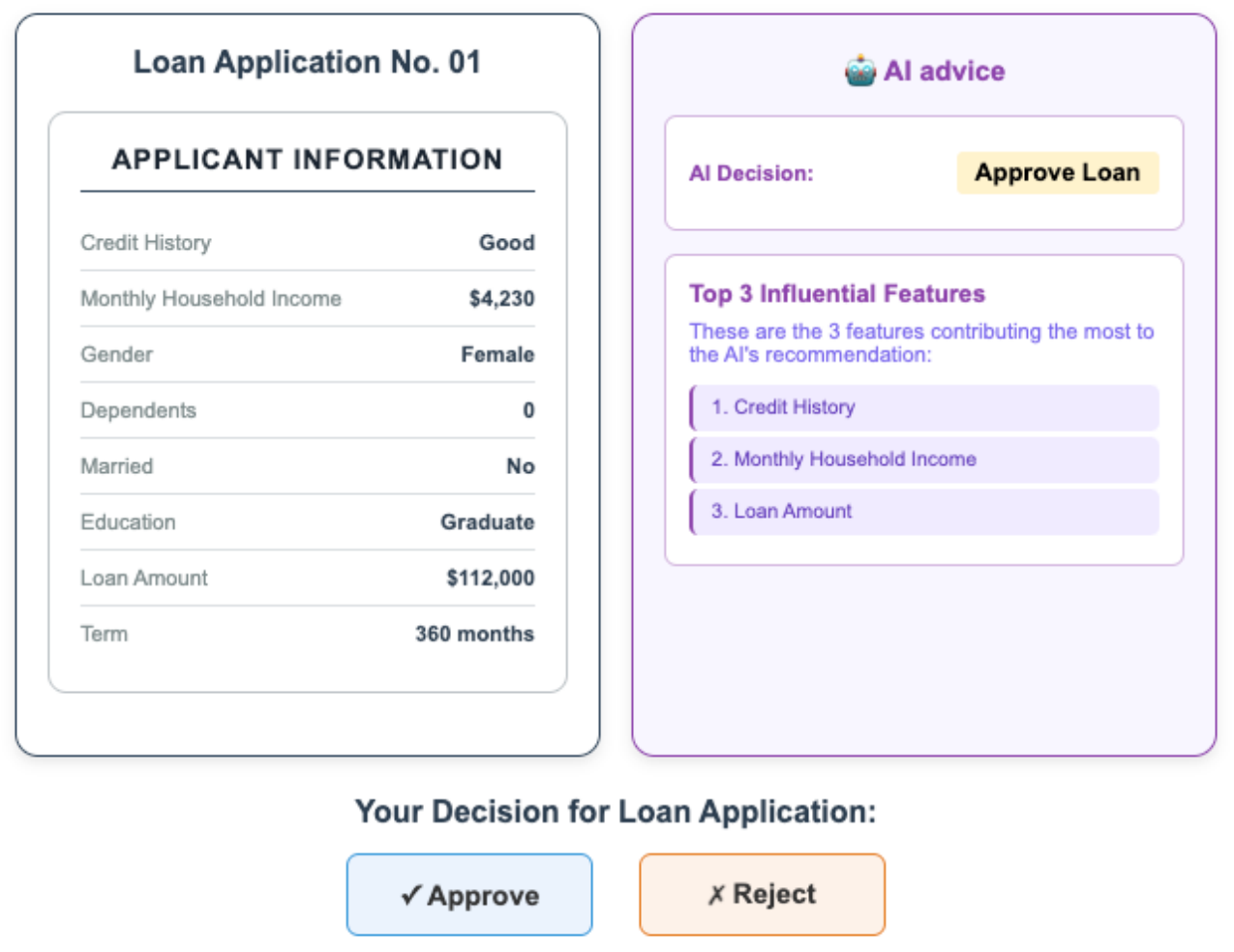}
        \caption{Loan decision interface showing applicant information and AI advice.}
        \Description{Screenshot of loan application interface showing applicant details in left panel (credit history: good, monthly income: \$4,230, gender: female, dependents: 0, married: no, education: graduate, loan amount: \$112,000, term: 360 months) and AI analysis panel on right showing "Approve Loan" decision with top 3 influential features listed as credit history, monthly household income, and loan amount. Bottom shows approve/reject decision buttons.}
        \label{fig:advice}
    \end{subfigure}
    \begin{subfigure}[b]{0.48\textwidth}
        \centering
        \includegraphics[width=\linewidth]{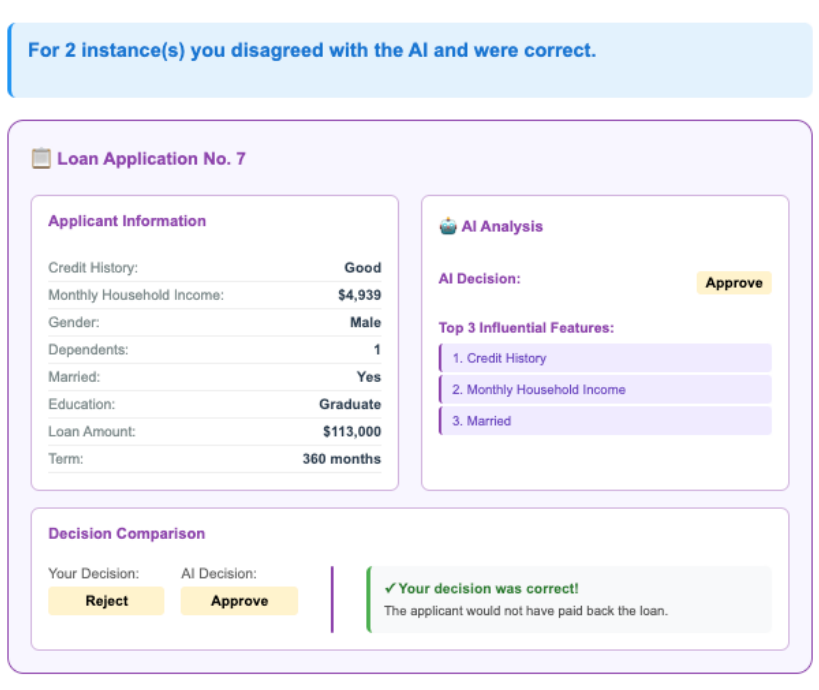}
        \caption{Exemplary feedback showing successful overrides of the AI advice.}
        \Description{Two loan application feedback examples. Application No. 7 shows a male applicant with $4,939 monthly income, 1 dependent, married, requesting $113,000. AI recommended approval, human chose rejection, outcome shows human was correct as the applicant would not have repaid. Application No. 10 shows a male applicant with $8,334 monthly income, 0 dependents, married, requesting $160,000. Similar pattern with AI recommending approval, human rejecting, and human decision proven correct.}
        \label{fig:feedback}
    \end{subfigure}
    \caption{Experimental interface for human-AI collaborative loan decisions. Panel (a) displays the decision-making screen where participants view applicant attributes alongside an AI recommendation and its top three influential features. Panel (b) illustrates the feedback on past overrides, which highlights cases where human override of the AI advice led to correct outcomes (successful condition).}
\label{fig:interface}
\end{figure}

\textbf{AI System and Instance Selection.} We split our dataset in 70\% training and 30\% test data. Using the training data, we train a Random Forest classifier that achieves 66.57\% accuracy on the test data. This moderate accuracy level was intentionally selected to ensure a sufficient frequency of AI errors and facilitate human-AI disagreements. For the experiment, we curate a balanced set of 48 instances, including 15 correct rejections, 15 correct approvals, 9 incorrect rejections, and 9 incorrect approvals, based on ground truth labels and AI predictions, mirroring overall AI accuracy with 62.5\% correct advice. For each participant, we randomly sample 18 instances from this curated set. \Cref{fig:advice} shows an example of how the AI advice was presented to participants. 

\textbf{Explanations.} Each AI recommendation came with an explanation that listed the three features contributing most to it according to SHAP \citep{Lundberg2017}. We presented a ranked list without numerical values to keep the explanations accessible to participants without a technical background \citep{cheng2019explaining}, and an informal pretest with ten people outside the study confirmed that the rankings conveyed the AI's reasoning. Since our focus is on feedback on past overrides, the explanation format was identical across conditions.

\textbf{Treatments.} We employ a between-subjects design with four conditions across three rounds. In the \textit{Control} condition, participants complete all 18 loan approval decisions without receiving any feedback about their performance or overrides of AI advice. The three feedback conditions provide systematic feedback on past overrides of AI advice after rounds 1 and 2, highlighting instances where participants disagreed with the AI advice and whether the final decision was correct or incorrect. Specifically, there are two unidirectional feedback conditions: in the \textit{Successful} condition, feedback shows those instances of the previous round where participants disagreed with the AI but made the correct decision according to ground truth; in the \textit{Unsuccessful} condition, feedback shows those instances of the previous round where participants disagreed with the AI but made incorrect decisions. The \textit{Balanced} condition combines both types of feedback, showing participants all instances of the previous round where participants successfully or unsuccessfully overrode the AI advice. Participants in each feedback condition were explicitly informed about the selective nature of their feedback during the introduction phase. We present all feedback on past overrides as a brief summary after the first and second rounds, showing, for each instance, the original case details, AI advice, SHAP-based explanation, participant decision, and ground truth. \Cref{fig:feedback} visualizes the feedback interfaces for the different Treatments. We restrict the feedback to overrides and thus to two of the four cases of the reliance taxonomy. An override is the only case in which the participant's judgment and the AI advice diverge, so its outcome shows which of the two was right and makes the participant's decision to disagree with the AI the object of reflection. Under adherence, the participant's decision coincides with the AI advice, and its outcome shows whether the AI was right but not whether the participant's own judgment would have differed. We discuss the consequences of omitting adherence outcomes in \Cref{sec:discussion-reliance}.

\begin{figure}
    \centering
    \includegraphics[width=1\linewidth]{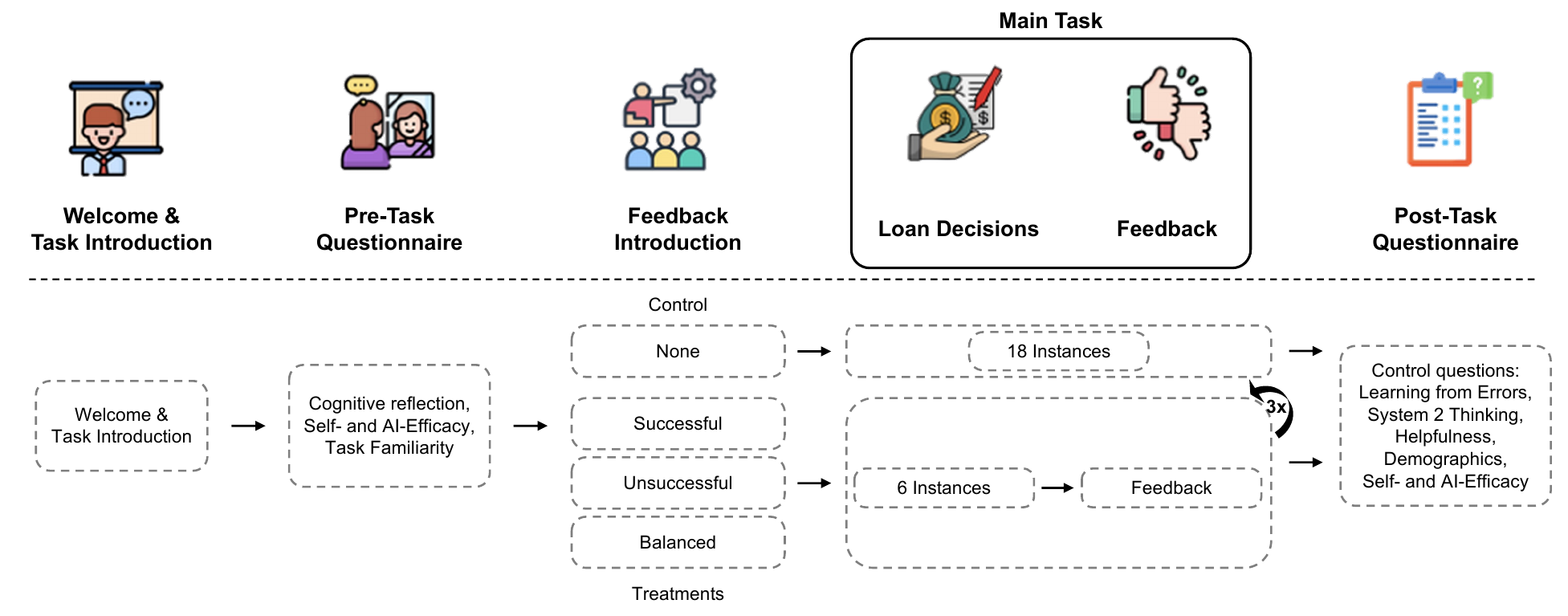}
    \caption{Experimental procedure showing the study flow from initial task introduction through the main decision-making phase to the post-task questionnaire, which repeats the efficacy measures.}
    \label{fig:procedure}
\end{figure}

\textbf{Study Procedure.} As shown in \Cref{fig:procedure}, the experiment starts with a welcome and task introduction where participants receive detailed instructions about the loan approval decision-making task. Participants then complete a cognitive reflection test to assess their tendency to engage in reflective thinking \citep{frederick2005cognitive}, followed by pre-task measurements of perceived self-efficacy and perceived AI efficacy \citep{spreitzer1995psychological} and task familiarity. Next, participants are randomly allocated to one of the four conditions and receive an introduction to the feedback interface (for participants in treatment conditions only). Following this, all participants work through a single practice example that demonstrates the interface, including the case details and AI advice, along with a SHAP-based explanation to ensure familiarity with the task. In addition, this practice example serves as a comprehension check. The main experimental phase consists of three rounds of six randomly selected loan approval decisions each. To ensure ecological validity, we delivered feedback in rounds to simulate the natural delay between a decision and its outcome.  Upon completion of all three rounds, participants are asked to complete a post-questionnaire. We distribute four attention checks across the study procedure, including one within the task instances and three within questionnaire items, all requiring participants to select specified responses. Feedback is generated from participants' own overrides, so participants who override the AI more often also receive more feedback (\Cref{app:feedback-frequency}).

\textbf{Measurement.} To assess how feedback on past overrides affects reflection, we measure reflection-in-action through system 2 thinking \citep{novak2009fit}, and we measure retrospective reflection-on-action through perceived learning from errors \citep{rybowiak1999error} in the post-task questionnaire. We measure both constructs after the task, since concurrent approaches such as think-aloud alter the behavior they are meant to record. We therefore selected the scale of \citet{novak2009fit}, which was developed for retrospective self-assessment of whole task episodes. Since both scales rest on participants' own perception, we complement them with two objective anchors, i.e., the cognitive reflection test before the task and page-level timestamps (see \Cref{app:metric-validation}). We operationalize reflection-on-action as learning from the outcomes of past decisions and measure it with the Error Orientation Questionnaire of \citet{rybowiak1999error}, an established instrument whose items are framed around recognizing and learning from errors. We retained the item wording apart from the reference to the task. Its error framing may understate reflection-on-action under feedback on successful overrides (Section~\ref{sec:limitations}).

To assess how feedback on past overrides affects reliance and human-AI team performance, we measure the accuracy of decisions by comparing participants' decisions and AI advice with ground truth labels from the dataset. We operationalize reliance as observed behavior based on \citet{schoeffer2024explanations}. Therefore, we compare a participant's final decision with the AI advice for every decision. This yields adherence when the two coincide and an override when they diverge. Crossing this with the correctness of the AI advice results in four cases: correct adherence, detrimental adherence, corrective overriding, and detrimental overriding. We report reliance as the share of decisions in which a participant adhered to the AI advice, and our feedback on past overrides draws on the two overriding cases of this taxonomy. We refer to the accuracy of the final decisions as human-AI team performance, following research on AI-assisted decision-making \citep{bansal2021does, schemmer2023appropriate}. The term describes the joint outcome of a human deciding with AI advice and does not presuppose that participants perceive the AI as a teammate. As control variables, we used the cognitive reflection test, perceived self-efficacy and AI efficacy, as well as task familiarity. Since the self-efficacy and AI-efficacy items are repeated after the task, we can also compute how both beliefs changed over the course of the task (see \Cref{sec:mediation-results}). Furthermore, we collected socio-demographic data in the post-task questionnaire as control variables. An overview of variables, including all items per construct, is provided in \Cref{tab:all_measures} in the Appendix.

\textbf{Qualitative Analysis.} Our qualitative analysis pursued two complementary aims: a confirmatory one, triangulating our quantitative findings, and an exploratory one, shedding light on the participants' reasoning and reflection processes. To that end, we asked open-ended questions in the post-task questionnaire addressing the \textit{decision cues} participants drew on and the \textit{effects} the feedback had on their beliefs and behavior, adapting questions from \citet{he2023knowing}. \Cref{tab:all_measures} in the Appendix lists the questions. We analyzed the responses using codebook thematic analysis \citep{braun2021onesize}. We developed a two-facet codebook through a hybrid deductive–inductive process \citep{fereday2006demonstrating}. Seed codes for the \textit{effects} facet were derived from the theoretical constructs underlying our experimental measures (reliance, AI-efficacy, self-efficacy, reflection-in-action, and reflection-on-action) and were then refined and extended inductively. The \textit{decision cues} facet was coded purely inductively, without a predefined template. Given the brevity of open-ended responses, we treated each participant's full response to a question as the unit of coding, allowing multiple codes per response; codes were then grouped into themes. Two authors independently coded a stratified subset of 108 responses (20\% of the corpus) using the final codebook. As responses can carry multiple themes, we assessed agreement with Krippendorff's multi-valued $\alpha$ ($mv\alpha$)~\citep{krippendorff2016reliability}, which extends Krippendorff's $\alpha$~\citep{krippendorff2018content} to units coded with sets of values. On the independent codings, agreement at the theme level was $mv\alpha = .80$. Per-theme agreement (binary Krippendorff's $\alpha$) ranged from .73 to .89. Disagreements were resolved through discussion. The remaining responses were coded in an LLM-assisted setup with Claude (Opus 5.5), where all final codes were verified by one author. We report inter-rater reliability between both human coders and the LLM-assisted coding using Krippendorff's multi-valued $\alpha$, yielding $mv\alpha_{Coder1-LLM} = .84$ and $mv\alpha_{Coder2-LLM} = .76$, respectively. We then used the coding in two complementary ways. First, we quantified code and theme prevalences for \textit{effects} \citep{chi1997quantifying} and compared them across conditions to examine whether the qualitative patterns converged with our quantitative findings. Second, we analyzed the themes qualitatively to surface patterns not anticipated by our hypotheses.

\textbf{Participants.} We recruited participants through Prolific.co who were fluent in English, US residents, and maintained a high approval rating on the platform. We excluded participants who failed more than one attention check, the comprehension check, or always followed the AI advice without deviation, indicating no task engagement (n=6). The final sample consisted of 244 participants equally distributed over the four conditions. Participants completed the study in a median time of 16.41 minutes and received a base payment of \$2.25 and a bonus payment of five cents per correct decision, resulting in an average payout of \$2.76. The final sample consists of 49.6\% male and 50.4\% female participants with an average age of 44.09 years. ANOVA tests confirmed that control variables, including socio-demographic variables and cognitive reflection, AI efficacy, and self-efficacy, were well-balanced across treatment conditions ($p >0.60$). Additionally, AI accuracy of the randomly sampled instances was balanced across conditions ($F(3,240)=0.30, p=0.82$), with mean accuracy ranging from 61.9\% to 63.5\%.
\section{Results}
\label{sec-results}

\Cref{sec:cognitive_processes} examines how feedback on past overrides affects reflection-on-action and reflection-in-action (\textbf{RQ\ref{rq1}}). \Cref{sec:behavioral-outcomes} turns to reliance and human-AI team performance (\textbf{RQ\ref{rq2}}), separates both by the correctness of the AI advice, and tests whether changes in perceived AI efficacy and self-efficacy mediate the reliance shifts. \Cref{sec:qualitative} reports the themes in participants' open-ended responses, and \Cref{sec:integration} uses them to explain results that the quantitative analysis leaves open.

\subsection{Reflective Processes: In-Action and On-Action}
\label{sec:cognitive_processes}
To understand how our feedback on past overrides influences participants' reflection (\textbf{RQ\ref{rq1}}), we first examine whether they successfully activate the intended cognitive process of retrospective reflection-on-action. Additionally, as the feedback on past overrides may also influence reflection during decision-making, we measure participants' reflection-in-action.

Since both constructs are measured in the post-task questionnaire, we first assess whether they capture distinct processes. They are uncorrelated in the control condition ($r = 0.171$, $p = 0.189$). System 2 thinking correlates with the cognitive reflection test taken before the task ($r = 0.147$, $p = 0.022$) and with the time spent on the decisions ($r = 0.198$, $p = 0.002$), whereas perceived learning from errors correlates with neither ($r = 0.045$, $p = 0.487$ and $r = 0.096$, $p = 0.134$). System 2 thinking is thus tied to the decision phase, in which reflection-in-action occurs (\Cref{app:metric-validation}).

For the analysis, we conduct two separate ANCOVA tests with treatment condition as the between-subjects factor and cognitive reflection degree, AI efficacy, and self-efficacy as covariates. 

\subsubsection{Reflection-on-Action}
\begin{figure}[h!]
    \centering
    \includegraphics[width=0.75\linewidth]{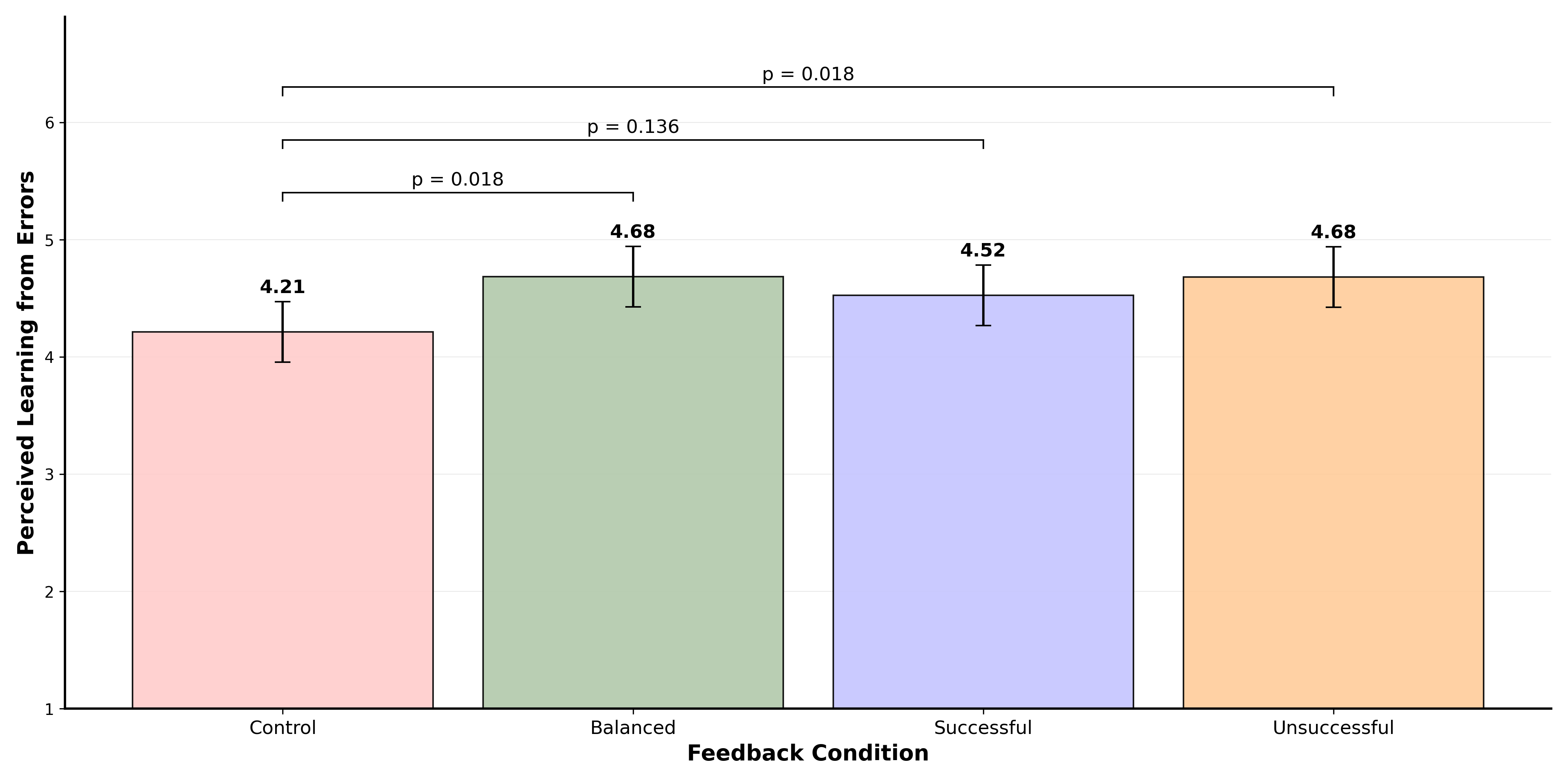}
    \caption{Perceived learning from errors (reflection-on-action) across conditions for adjusted means with 95\% confidence intervals.}
    \label{fig:learning_errors}
    \Description{Vertical bar chart of perceived learning from errors, the measure of reflection-on-action, by feedback condition. The y-axis spans the full 6-point response scale from 1 to 6. Bars show adjusted means with 95 percent confidence intervals: Control 4.21, Balanced 4.68, Successful 4.52, and Unsuccessful 4.68. Significance brackets above the bars report each treatment against control, with p equals 0.018 for Balanced, p equals 0.136 for Successful, and p equals 0.018 for Unsuccessful. Balanced and Unsuccessful feedback raise perceived reflection-on-action significantly above control, while Successful feedback does not.}
\end{figure}

The analysis indicates that feedback on past overrides increases perceived learning from errors. The ANCOVA reveals a significant main effect of treatment condition ($F(3,237) = 2.85, p = 0.038$). Perceived AI efficacy emerges as a highly significant covariate ($p < 0.001$), while cognitive reflection ($p = 0.490$) and self-efficacy ($p = 0.299$) show no significant effects. \Cref{fig:learning_errors} visualizes the adjusted means across conditions, and \Cref{tab:learning_errors} in the Appendix reports the detailed results.

The analysis of adjusted means (\Cref{tab:learning_errors} in the Appendix) shows that all three forms of feedback on past overrides show higher adjusted reflection-on-action compared to the control condition ($\mu = 4.21$). The balanced feedback condition ($\mu = 4.68$) and the unsuccessful feedback condition ($\mu = 4.68$) show equally strong effects, while the increase under successful feedback is smaller and not significant ($\mu = 4.52$).

Following our hypothesis \ref{hyp1}, we perform one-sided contrast tests to examine whether treatments increase retrospective reflection. To correct for family-wise errors, we use Sidak adjustments. The results confirm significant improvements for two conditions: the balanced feedback condition shows a significant increase over control ($\Delta = 0.471$, $p = 0.018$, $d = 0.46$), as does the unsuccessful feedback condition ($\Delta = 0.468$, $p = 0.018$, $d = 0.46$). The successful feedback condition, while not reaching statistical significance over the control condition, shows a small effect ($\Delta = 0.311$, $p = 0.136$, $d = 0.30$). With the balanced and the unsuccessful feedback reaching significance, we find partial support for H\ref{hyp1}.

\subsubsection{Reflection-in-Action}
Treatment condition also affects System 2 thinking ($F(3,237) = 3.39, p = 0.019$). Cognitive reflection ($p = 0.015$) and self-efficacy ($p = 0.019$) emerge as significant covariates, while perceived AI efficacy showed a marginally significant effect ($p = 0.079$). \Cref{fig:system_2} visualizes the adjusted means across conditions and \Cref{tab:system2_thinking} in the Appendix reports the detailed results.

\begin{figure}[h!]
    \centering
    \includegraphics[width=0.75\linewidth]{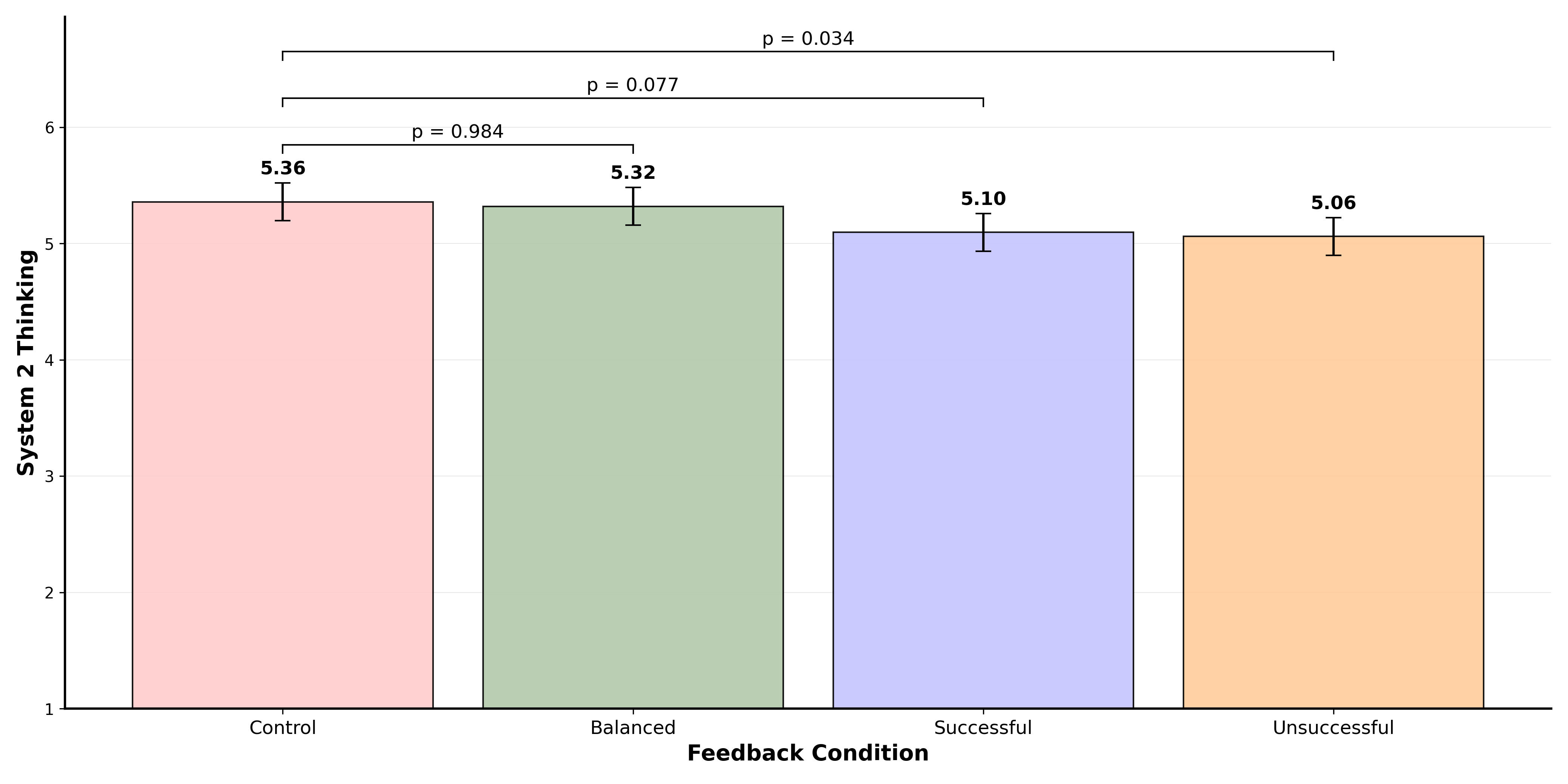}
    \caption{System 2 thinking (reflection-in-action) across conditions for adjusted means with 95\% confidence intervals.}
    \label{fig:system_2}
    \Description{Vertical bar chart of System 2 thinking, the measure of reflection-in-action, by feedback condition. The y-axis spans the full 6-point response scale from 1 to 6. Bars show adjusted means with 95 percent confidence intervals: Control 5.36, Balanced 5.32, Successful 5.10, and Unsuccessful 5.06. Significance brackets above the bars report each treatment against control, with p equals 0.984 for Balanced, p equals 0.077 for Successful, and p equals 0.034 for Unsuccessful. Unsuccessful feedback lowers perceived reflection-in-action significantly relative to control, Successful feedback shows a marginal reduction, and Balanced feedback is indistinguishable from control.}
\end{figure}

We find that both unidirectional feedback conditions show lower adjusted reflection-in-action relative to our control condition ($\mu = 5.36$). The unsuccessful condition shows the largest decrease ($\mu = 5.06$), followed by the successful condition ($\mu = 5.10$). The balanced feedback condition preserves perceived reflection-in-action at the level of control ($\mu = 5.32$).

Two-sided contrast analyses using Sidak-adjusted comparisons reveal that participants receiving unsuccessful feedback demonstrate significantly lower reflection-in-action compared to control ($\Delta = -0.297$, $p = 0.034$, $d = 0.46$), while the successful condition exhibits a marginally significant reduction ($\Delta = -0.262$, $p = 0.077$, $d = 0.41$). Conversely, balanced feedback participants show no significant change from baseline processing ($\Delta = -0.038$, $p = 0.984$, $d = 0.06$).

These findings collectively show a tension in unidirectional feedback design: while balanced and unsuccessful feedback increase retrospective reflection-on-action, unidirectional feedback is also associated with lower reflection-in-action (significant for unsuccessful, marginal for successful feedback). Balanced feedback does not show this cognitive trade-off, increasing retrospective reflection while maintaining concurrent reflection.

\needspace{14\baselineskip}
\subsection{Behavioral Outcomes: Reliance Patterns and Human-AI Team Performance}
\label{sec:behavioral-outcomes}

We now turn to reliance behavior and decision-making performance (\textbf{RQ\ref{rq2}}).

\subsubsection{Reliance Outcomes During the Feedback Rounds}
\label{sec:reliance-outcomes}
Before testing the treatment effects, we describe how participants' decisions in the two feedback rounds distribute over the four reliance outcomes of \citet{schoeffer2024explanations}, since these outcomes determine what the feedback on past overrides shows. Participants followed correct AI advice in 50.5\% of decisions and incorrect AI advice in 22.0\%, and they overrode the AI successfully in 14.6\% and unsuccessfully in 12.9\% of decisions. Following incorrect advice was thus the most frequent error, and 94.7\% of participants made it at least once. Overrides, the cases our feedback covers, make up 27.5\% of decisions, on average 3.3 cases per participant across both feedback rounds. All errors of the human-AI team, that is, unsuccessful overrides and adherence to incorrect advice, make up 34.9\% of decisions, and for 49.5\% together with successful overrides.

\subsubsection{Reliance on AI Advice} 

To analyze how our feedback on past overrides influences reliance, we analyze participants' reliance behavior in round 3.  Mixed-effects logistic regressions across rounds yield no significant Treatment $\times$ Round interaction, both with covariates ($\chi^2(3) = 5.45$, $p = 0.142$) and without ($\chi^2(6) = 9.88$, $p = 0.130$). \Cref{app:longitudinal} reports both models alongside the round 2 ANCOVA. We employ an ANCOVA test with reliance on AI advice in round 3 as the target variable, keeping cognitive reflection degree, AI efficacy, and self-efficacy as covariates. To account for participants' baseline reliance, we further include their reliance from round 1, which differs across conditions ($F(3,240) = 3.09$, $p = 0.028$) despite balanced pre-task covariates. Additionally, we include feedback exposure as two covariates: the number of feedback instances each participant received after round 1 and after round 2. 

The ANCOVA reveals a significant main effect of treatment condition ($F(3,234) =$ $4.09$, $p=0.007$).  Feedback volume is a consequence of participants' own override behavior and differs across conditions ($p < 0.001$). Within the three feedback conditions, adding the exact number of feedback instances shown to each participant leaves the differences between conditions unchanged ($F(2,176) = 6.26$ without and $F(2,175) = 6.30$ with the count, both $p = 0.002$), and feedback count predicts neither reliance ($p = 0.629$) nor reflection-on-action ($p = 0.236$), reflection-in-action ($p = 0.962$), or performance ($p = 0.113$). \Cref{app:feedback-frequency} reports this check.

\begin{figure}[h!]
    \centering
    \includegraphics[width=0.75\linewidth]{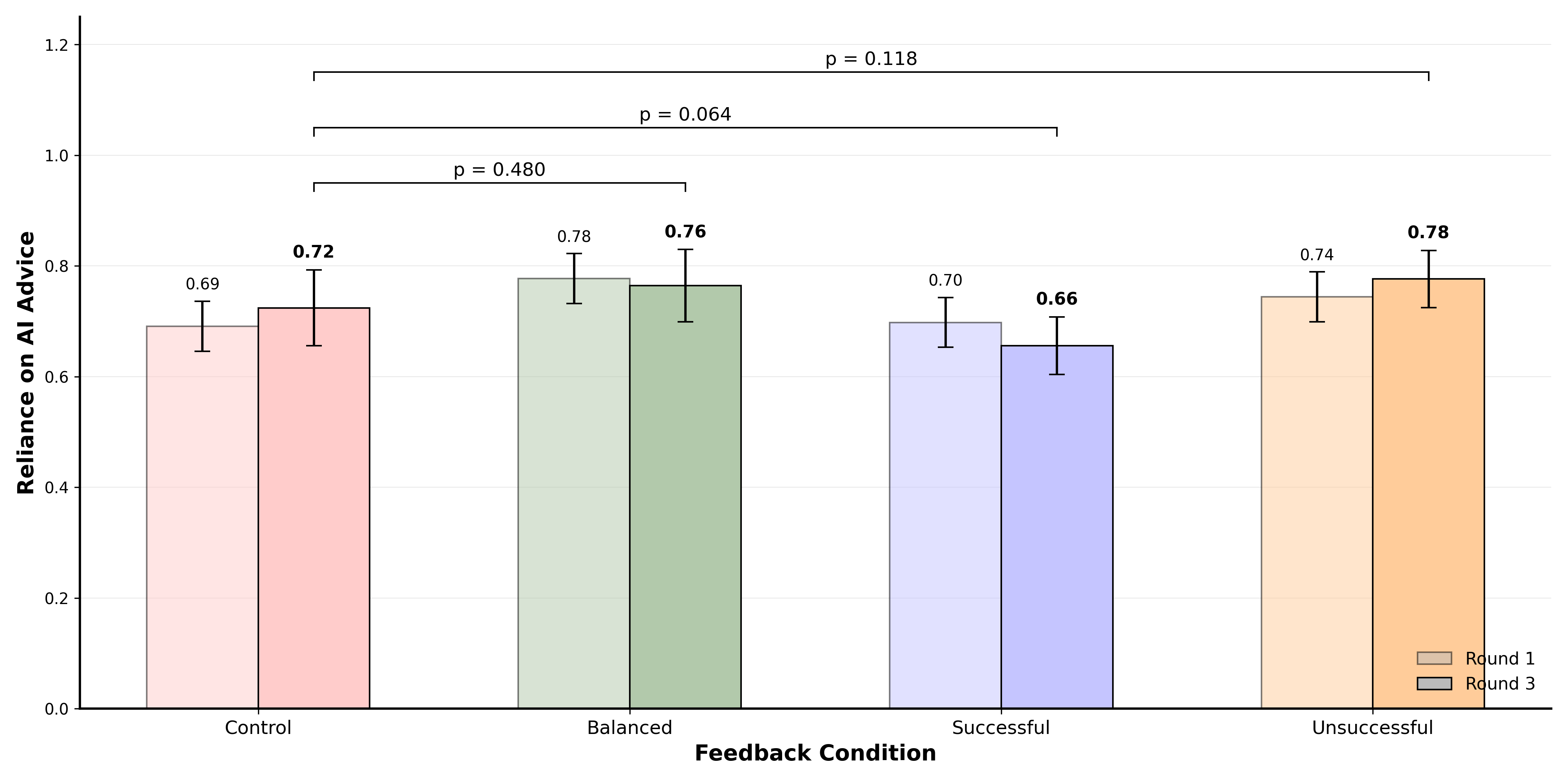}
    \caption{Adjusted means of adherence to AI advice in round 1 (lighter bars) and in the final round (after seeing feedback on past overrides) across conditions, with 95\% confidence intervals.}
    \label{fig:reliance_round3}
    \Description{Vertical bar chart of reliance on AI advice by feedback condition, with paired bars for round 1 (lighter) and round 3 (solid) in each condition, all showing adjusted means with 95 percent confidence intervals. The y-axis runs from 0 to 1.25. Control moves from 0.69 to 0.72, Balanced from 0.78 to 0.76, Successful from 0.70 to 0.66, and Unsuccessful from 0.74 to 0.78. Significance brackets compare each treatment against control in round 3, with p equals 0.480 for Balanced (two-sided), p equals 0.064 for Successful, and p equals 0.118 for Unsuccessful (both one-sided). No comparison reaches significance; Successful feedback shows the lowest adherence.}
\end{figure}

Comparing adjusted means of reliance across treatments yields the following ordering (\Cref{tab:reliance_round3} in the Appendix). The unsuccessful feedback condition produces the highest reliance ($\mu =$ $0.776$), followed by the balanced feedback condition ($\mu =$ $0.764$). The control condition maintains moderate adherence levels ($\mu =$ $0.724$), while the successful feedback condition demonstrates the lowest adherence to AI advice ($\mu = 0.656$) as visualized in \Cref{fig:reliance_round3}.

To test our second hypothesis, we perform one-sided tests for our unidirectional feedback conditions (unsuccessful and successful; here we expect directional effects, see H\ref{hyp2}) and a two-sided test for our balanced feedback condition (here we do not assume directional effects, see H\ref{hyp2}). The results show marginal significance for our successful condition ($p =$ $0.064$) and no significance for our unsuccessful condition ($p =$ $0.118$) and balanced condition ($p =$ $0.480$). The effect sizes point in the hypothesized directions. The successful condition shows a small effect on reduced adherence compared to the control condition ($d = 0.33$), while the unsuccessful condition shows a small effect on increased adherence ($d = 0.26$). The balanced feedback condition shows a small effect toward higher adherence ($d =$ $0.20$). The two unidirectional conditions differ from each other more than either differs from control, by a medium effect size ($d = 0.59$). Thus, we find partial support for our second hypothesis H\ref{hyp2}.

\subsubsection{Human-AI Team Performance}

We use ANCOVA tests to analyze performance in the final round (round 3). Similar to our analysis of reliance behavior, we include cognitive reflection degree, AI, and self-efficacy as well as the baseline performance of round 1---to account for performance outcomes before any feedback on past overrides---and the amount of feedback exposure during the experiment as covariates. The ANCOVA reveals a marginal main effect of treatment conditions ($F(3,234) =$ $2.15$, $p =$ $0.095$). 

\begin{figure}[h!]
    \centering
    \includegraphics[width=0.75\linewidth]{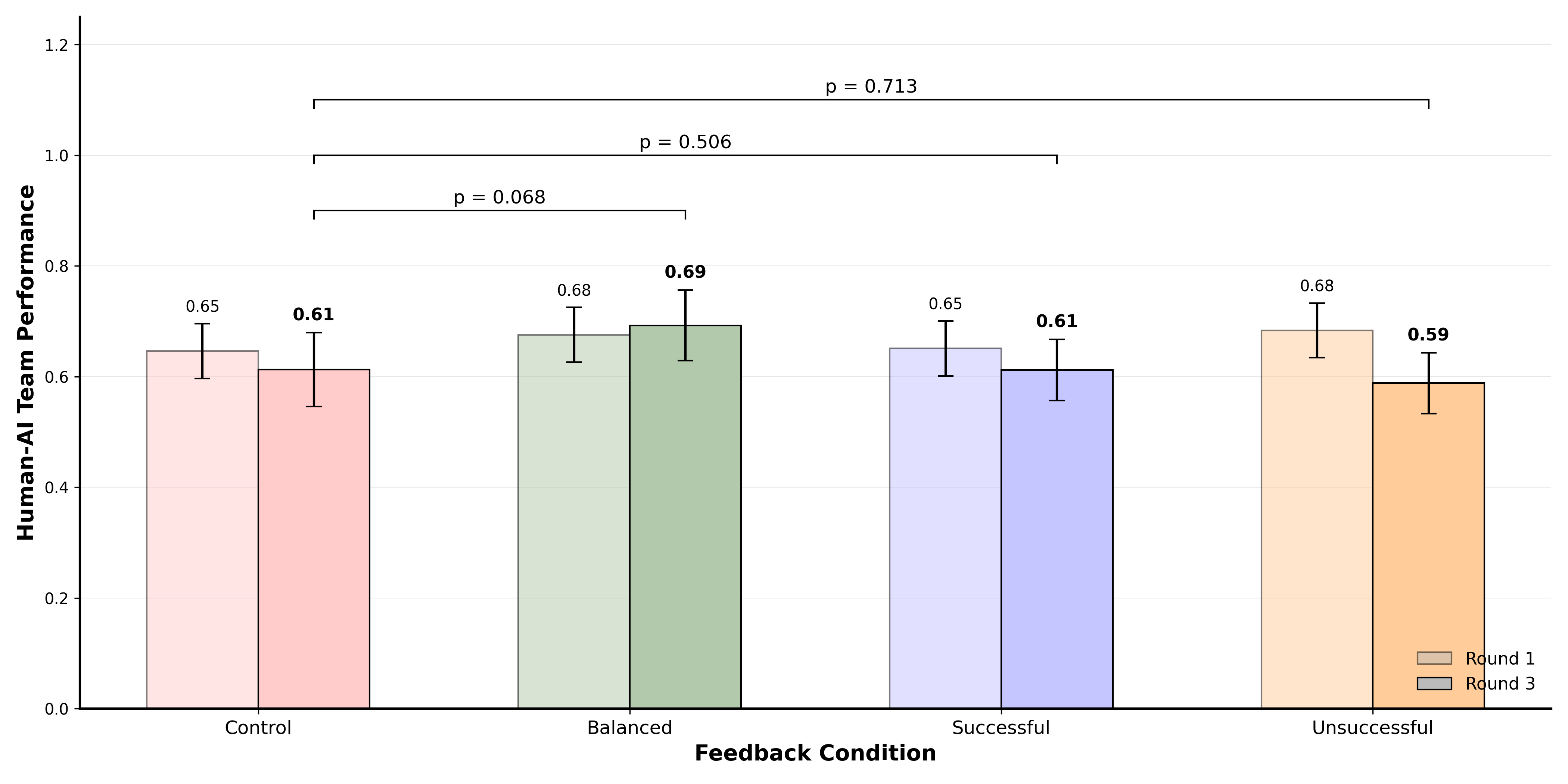}
    \caption{Adjusted means of human-AI team performance in round 1 (lighter bars) and in the final round (after seeing feedback on past overrides) across conditions, with 95\% confidence intervals.}
    \label{fig:performance_round3}
    \Description{Vertical bar chart of human-AI team performance by feedback condition, with paired bars for round 1 (lighter) and round 3 (solid) in each condition, all showing adjusted means with 95 percent confidence intervals. The y-axis runs from 0 to 1.25. Control moves from 0.65 to 0.61, Balanced from 0.68 to 0.69, Successful from 0.65 to 0.61, and Unsuccessful from 0.68 to 0.59. Significance brackets compare each treatment against control in round 3 with one-sided tests, with p equals 0.068 for Balanced, p equals 0.506 for Successful, and p equals 0.713 for Unsuccessful. No comparison reaches significance.}
\end{figure}

The analysis of adjusted means (\Cref{tab:performance_round3} in the Appendix) yields the ordering visualized in \Cref{fig:performance_round3}. The balanced feedback condition achieves the highest performance ($\mu = 0.692$), followed by the control baseline ($\mu =$ $0.613$). The successful condition maintains performance similar to control ($\mu = 0.612$), while the unsuccessful condition shows the lowest performance ($\mu =$ $0.588$).

To test H\ref{hyp3}, we use a one-sided contrast test for comparing our different treatments to our control condition. Although these tests do not reach statistical significance (balanced $p = 0.068$, successful $p = 0.506$, unsuccessful $p = 0.713$), the conditions differ in how far they move from control. The balanced feedback condition shows the largest difference from control, which is marginally significant ($d =$ $0.37$), while unidirectional feedback conditions only show negligible effects ($d =$ $0.00$, $d =$ $0.11$).  Consequently, we find no support for our final hypothesis H\ref{hyp3}.

\subsubsection{Context-Dependent Effects} To better understand observed treatment effects, we further analyzed reliance and performance separately for instances in which AI advice was correct versus incorrect.

In this context, reliance and performance measures become equivalent indicators of appropriate collaborative behavior. Specifically, when AI advice is correct, optimal performance requires adherence to AI advice, whereas when AI advice is incorrect, optimal performance requires overriding AI advice. 

For this analysis, we focus on individual instances rather than previous aggregate measures at the participant level, which allows us to subset the data based on whether AI advice was correct or incorrect and to employ mixed-effects logistic regression models. Since none of the covariates reached significance in the round 3 models of reliance and performance (\Cref{tab:reliance_round3,tab:performance_round3}), we proceed with treatment condition as the sole fixed effect and include random intercepts for participants and instances. As previously, we focus our analysis on round 3 after full feedback exposure. 

\begin{table}[h!]
\centering
\caption{Mixed-Effects Logistic Regression: Performance by AI Advice Correctness}
\label{tab:reliance_merged}
\begin{tabular}{l*{4}{c}*{4}{c}}
\toprule
& \multicolumn{4}{c}{\textbf{AI Advice Correct}} & \multicolumn{4}{c}{\textbf{AI Advice Incorrect}} \\
\cmidrule(lr){2-5} \cmidrule(lr){6-9}
\textbf{Treatment} & \textbf{Coeff.} & \textbf{SE} & \textbf{z-value} & \textbf{p-value} & \textbf{Coeff.} & \textbf{SE} & \textbf{z-value} & \textbf{p-value} \\
\midrule
(Intercept) & 2.03 & 0.41 & 4.9 & \textbf{$<$ 0.001} & -0.78 & 0.55 & -1.43 & 0.153 \\
Balanced & 0.61 & 0.40 & 1.55 & 0.122 & -0.16 & 0.39 & -0.42 & 0.67 \\
Successful & -0.62 & 0.37 & -1.67 & 0.095 & 0.26 & 0.39 & 0.69 & 0.48 \\
Unsuccessful & 0.32 & 0.40 & 0.81 & 0.416 & -0.83 & 0.39 & -2.15 & \textbf{0.03} \\
\bottomrule
\end{tabular}
\begin{tablenotes}
\item AI Correct: N = 896 observations, 243 participants, 30 instances. Random effects: participants ($\sigma^2$ = 1.421) and instances ($\sigma^2$ = 2.46). Model significance: $\chi^2$(3) = 11.473, p = 0.009.
\item AI Incorrect: N = 568 observations, 234 participants, 18 instances. Random effects: participants ($\sigma^2$ = 0.89) and instances ($\sigma^2$ = 4.03). Model significance: $\chi^2$(3) = 8.65, p = 0.034.
\item Each model contains only participants who saw at least one instance of the respective type in round 3, which comprises six decisions per participant. One participant saw only instances with incorrect AI advice and ten saw only instances with correct AI advice. The sample therefore remains 244 participants, and the 1,464 round 3 decisions (896 + 568) are split completely between the two models.
\end{tablenotes}
\end{table}

\textbf{When AI Advice is Correct.} The analysis of reliance when AI provided correct advice reveals significant treatment effects as demonstrated by the likelihood ratio tests ($\chi^2(3) = 11.473$, $p =$ $0.009$), indicating that feedback on past overrides influences participants' acceptance of correct AI advice.
The balanced feedback condition shows a non-significant difference in the direction of higher adherence to correct AI advice ($\beta = 0.61, p = 0.122$). In contrast, the successful feedback condition shows a marginally significant reduction in adherence to correct AI advice ($\beta = -0.62, p = 0.095$), resulting in reduced performance through inappropriate overriding of correct AI advice. The unsuccessful feedback condition shows no significant difference from control ($\beta = 0.32, p = 0.416$).

\textbf{When AI Advice is Incorrect.} For cases where AI provided incorrect advice, treatment effects are significant ($\chi^2(3) = 8.65, p = 0.034$).
The unsuccessful feedback condition shows significantly greater adherence to incorrect AI advice ($\beta = -0.83, p = 0.03$), resulting in impaired human-AI team performance. Both the balanced feedback condition ($\beta = -0.16,  p = 0.67$) and successful condition ($\beta = 0.26, p = 0.48$) show no significant effects.

Together, these findings indicate that our feedback on past overrides shapes human behavior in collaborative settings, with reliance behavior mirroring the reflection patterns observed. Balanced feedback shows the highest adjusted human-AI team performance, marginally above control. The context-dependent analyses point to why the global differences remain small, since opposing effects for correct and incorrect AI advice partially cancel each other out. Feedback on successful overrides can thus lead to overriding correct AI advice in later decisions (marginally), and feedback on unsuccessful overrides to adhering to incorrect AI advice, whereas balanced feedback shows neither pattern.

\subsubsection{Mechanisms Underlying the Reliance Shifts}
\label{sec:mediation-results}
The opposing reliance shifts may follow from participants revising their beliefs about the AI or their own, or from the feedback acting on behavior without such a revision. To distinguish both explanations, we conduct a subgroup analysis and compute delta scores for perceived AI efficacy and perceived self-efficacy (post-task minus pre-task) and estimate causal mediation models with 1,000 quasi-Bayesian simulations for each feedback condition against control, with reliance in round 3 as the outcome. Both constructs were measured before the first round and again in the post-task questionnaire after round 3. The mediator is therefore not measured before the outcome, and the models test whether the reliance shifts are consistent with a revision of these beliefs without establishing temporal precedence. We report each mediator in turn, covering all three conditions, and \Cref{tab:mediation} in the Appendix gives the full models.

\textbf{Change in perceived AI efficacy.} Perceived AI efficacy declines over the task in the control condition and declines less or increases in every feedback condition, most strongly under unsuccessful feedback ($a = 0.127$, $SE = 0.031$) and least under balanced feedback, where the difference from control is not distinguishable from zero ($a = 0.046$, $SE = 0.032$), and participants whose perception of the AI improves rely on it more ($b = 0.315$, $SE = 0.099$). In the unsuccessful condition, this indirect path explains the entire association with reliance, since the mediated effect is significant (ACME $= 0.040$, $p < 0.001$) while the direct effect is not (ADE $= 0.016$, $p = 0.648$). The total effect remains non-significant ($0.056$, $p = 0.108$), so the mediated path characterizes the mechanism without decomposing an established total effect. In the successful condition, the two paths run in opposite directions. The direct effect reduces reliance (ADE $= -0.097$, $p = 0.006$) while the mediated path increases it (ACME $= 0.024$, $p = 0.038$), which holds the total effect at the threshold of significance ($-0.073$, $p = 0.052$). In the balanced condition, neither path reaches significance.

\textbf{Change in perceived self-efficacy.} No condition shifts perceived self-efficacy, and no mediated effect through it approaches significance (ACME $= -0.002$, $p = 0.716$ in the successful condition, $p > 0.54$ in the other two). The direct effect of successful feedback remains significant when self-efficacy is included (ADE $= -0.073$, $p = 0.044$). Its total effect is simulated within each mediation model and therefore differs slightly between the two models ($-0.075$, $p = 0.036$, here and $-0.073$, $p = 0.052$, with change in perceived AI efficacy as mediator). We treat it as marginal, in line with the round 3 contrast against control ($p = 0.064$). The reduced adherence under successful feedback is therefore not an expression of better perception of one's own abilities.

Change in perceived AI efficacy plays a different role in each of the two unidirectional conditions. Increased adherence following the unsuccessful feedback intervention is consistent with better perception of the AI's abilities. Under successful feedback, perceived AI efficacy declines less over the task than in the control condition. Since participants with a more positive change in perceived AI efficacy adhere more to the AI advice, this smaller decline alone would increase adherence. Adherence nevertheless decreases, so the change in perceived AI efficacy cannot explain the reduction. \Cref{app:mediation} reports the underlying mediator and outcome models.

\subsection{Qualitative Findings: Decision Cues and Effects}
\label{sec:qualitative}

To complement our quantitative results, we analyze participants' open-ended responses on the influence of the provided feedback on past overrides (RQ1, RQ2). Our codebook thematic analysis focuses on two facets: decision cues participants drew on and the effects the feedback had on their beliefs and behavior. The \textit{decision cues} facet comprises four themes: the outcomes the feedback displayed, the individual instances it presented, the output of the AI system, and the details of the loan application itself. The \textit{effects} facet comprises eight themes spanning adjustments of reliance behavior, beliefs about the AI and about themselves, reflection-in-action, three perspectives of reflection-on-action, as well as statements of marginal influence and remarks on the feedback itself. Since participants in the control condition received no feedback on past overrides and accordingly reported little in response to these questions, the following analysis covers the three feedback conditions.

\Cref{tab:prevalence-themes} reports the prevalence of each theme across the three feedback conditions, and \Cref{tab:prevalence-full} in Appendix~\ref{app:qual} reports the prevalence of each code. Responding to open-ended questions, participants confirm their engagement with the displayed feedback interventions and their adjustments of reliance behavior as well as perceptions of the AI and themselves. The qualitative analysis also surfaces three insights that complement our initial assumptions on how users respond to feedback on past overrides.

\begin{table}[t]
\centering
\small
\caption{Prevalence of themes across feedback conditions. Cells report the percentage of participants whose open-ended responses carried the theme at least once. \Cref{tab:prevalence-full} in Appendix~\ref{app:qual} reports the prevalence of each code.}
\label{tab:prevalence-themes}
\begin{tabular}{@{}lrrrr@{}}
\toprule
Theme & Balanced & Successful & Unsuccessful & All \\
 & ($n=61$) & ($n=61$) & ($n=61$) & ($n=183$) \\
\midrule
\multicolumn{5}{@{}l}{\textit{Facet 1: Decision cues}} \\
Feedback: Outcomes & 45.9 & 49.2 & 47.5 & 47.5 \\
Feedback: Instances & 47.5 & 42.6 & 55.7 & 48.6 \\
AI Output & 31.1 & 36.1 & 29.5 & 32.2 \\
Task details & 50.8 & 63.9 & 65.6 & 60.1 \\
\midrule
\multicolumn{5}{@{}l}{\textit{Facet 2: Effects}} \\
Reliance behavior changed & 13.1 & 13.1 & 16.4 & 14.2 \\
Perceptions of AI and self changed & 29.5 & 41.0 & 29.5 & 33.3 \\
Reflection on Action: Task understanding improved & 50.8 & 36.1 & 63.9 & 50.3 \\
Reflection on Action: AI understanding improved & 26.2 & 29.5 & 26.2 & 27.3 \\
Reflection on Action: Strategy developed & 29.5 & 31.1 & 36.1 & 32.2 \\
Reflection in Action: Task approached more deliberately & 18.0 & 26.2 & 16.4 & 20.2 \\
Marginally influenced & 24.6 & 31.1 & 39.3 & 31.7 \\
Feedback expected to be more elaborate & 4.9 & 3.3 & 9.8 & 6.0 \\
\bottomrule
\end{tabular}
\Description{Table of theme prevalences in the three feedback conditions, balanced, successful, and unsuccessful, each with 61 participants, and across all 183 participants. Decision cues: feedback outcomes 45.9, 49.2, 47.5, and 47.5 percent; feedback instances 47.5, 42.6, 55.7, and 48.6 percent; AI output 31.1, 36.1, 29.5, and 32.2 percent; task details 50.8, 63.9, 65.6, and 60.1 percent. Effects: reliance behavior changed 13.1, 13.1, 16.4, and 14.2 percent; perceptions of AI and self changed 29.5, 41.0, 29.5, and 33.3 percent; task understanding improved 50.8, 36.1, 63.9, and 50.3 percent; AI understanding improved 26.2, 29.5, 26.2, and 27.3 percent; strategy developed 29.5, 31.1, 36.1, and 32.2 percent; task approached more deliberately 18.0, 26.2, 16.4, and 20.2 percent; marginally influenced 24.6, 31.1, 39.3, and 31.7 percent; feedback expected to be more elaborate 4.9, 3.3, 9.8, and 6.0 percent.}
\end{table}

\textit{Participants report engaging more often with feedback where they were wrong compared to where the AI was wrong.}
While participants across all three feedback conditions refer to the outcomes their feedback displayed, those receiving feedback on successful overrides (i.e., those where they were right) least often report engaging with the individual feedback instances that were shown to them. In contrast, those receiving feedback on unsuccessful overrides most often report engagement with the feedback instances.  In the balanced condition, participants mostly referred to those particular instances they had gotten wrong rather than to those in which the AI had erred (\textit{``I looked much more closely at the questions that I got incorrect [...] to try and figure out what made me make the wrong judgement''}, P1510). This suggests that, in a human-AI collaboration scenario, feedback confirming one’s own judgment leads to less reflection-on-action than feedback that contradicts it. This helps explain our quantitative observation that the successful condition did not reach significance in activating reflection-on-action.

\textit{The participants' responses indicate that reflection-on-action goes beyond mistakes.}
Our quantitative results show that reflection-on-action increases significantly only in the balanced and unsuccessful conditions, that is, when participants receive feedback on overrides in which they were wrong (Section~\ref{sec:cognitive_processes}). Following \citet{rybowiak1999error}, we measured reflection-on-action as learning from errors, capturing how participants use their errors to improve their work on the task. The qualitative responses support this finding. Participants receiving feedback on unsuccessful overrides most frequently describe an improved understanding of the task, revisiting why a case was misjudged and reconsidering how much weight particular applicant attributes deserve (\textit{``I tried to look at the cases where I was wrong and figure out what factors I was under valuing or over valuing''}, P1394). Correspondingly, improved task understanding is most prevalent in the balanced and unsuccessful conditions. However, the responses also suggest that reflection-on-action is not limited to correcting individual mistakes. Participants additionally describe an improved understanding of the AI (\textit{``It allowed me to understand the formula that the AI process was using''}, P1483) and the development of their decision strategy (\textit{``Seeing specific cases was correct especially in borderline scenarios I had [initially] declined made me recalibrate my risk assessment''}, P1444). These findings indicate that, in human-AI collaboration, reflection-on-action extends beyond correcting individual mistakes to understanding how the AI arrives at its recommendations and how participants approach their decisions more generally.

\textit{When shown their own errors, participants revise their view of the AI more readily than their view of themselves.}
Decreased perceptions of the AI's efficacy are reported most frequently in the successful condition (\textit{``[...] the AI was not always correct and couldn[`]t be fully relied on''}, P1451), and increased perceptions most frequently in the unsuccessful condition (\textit{``The information made me have more confidence in the AI's decisions''}, P1414). This symmetry follows from the feedback, since feedback on successful overrides displays cases in which the AI was wrong and the human was correct, whereas feedback on unsuccessful overrides displays cases in which the AI was correct and the human was wrong. Perceived self-efficacy shows no such symmetry as participants frequently report increased self-efficacy after successful overrides, but reports of decreased self-efficacy remain rare in all conditions, including the unsuccessful condition. A stable self-efficacy, however, does not imply growing confidence in one's own judgment, because participants may come to distrust the AI for particular cases without trusting themselves more. These findings complement our quantitative results, in which only change in perceived AI efficacy mediates the effect of feedback on reliance (\Cref{sec:mediation-results}).
\subsection{Integrating Quantitative and Qualitative Findings}
\label{sec:integration}

We now compare the quantitative and qualitative findings through a joint display. For our hypotheses on reflection, behavior and team performance, as well as perceptions, \Cref{tab:joint-display} sets the quantitative result beside the qualitative result, classifies their relation as confirmation, expansion, or discordance, and states the meta-inference that both support together \citep{venkatesh2013bridging}. The quantitative results compare each feedback condition against control, whereas the qualitative results compare only the three feedback conditions, since participants in the control condition had no feedback to report on.

\begin{table}[h!]
\centering
\footnotesize
\caption{Joint display of quantitative and qualitative findings. Relations are classified as confirmation, expansion, or discordance. Percentages refer to participants whose open-ended responses carry the respective theme or code (\Cref{tab:prevalence-full}).}
\label{tab:joint-display}
\begin{tabularx}{\textwidth}{@{}p{1.6cm}XXp{1.7cm}X@{}}
\toprule
\textbf{Dimension} & \textbf{Quantitative result} & \textbf{Qualitative result} & \textbf{Relation} & \textbf{Meta-inference} \\
\midrule
Reflection & Reflection-on-action higher than control under balanced and unsuccessful feedback (both $p = 0.018$), not under successful feedback ($p = 0.136$). Reflection-in-action lower under unsuccessful ($p = 0.034$) and, marginally, under successful feedback ($p = 0.077$) & Improved task understanding most frequent after unsuccessful (63.9\%) and least frequent after successful overrides (36.1\%). A more deliberate approach was reported most often after successful overrides (26.2\%), described as weighing one's own reasoning against the AI as often as analyzing cases more carefully, and not more often after unsuccessful overrides (16.4\%) than after balanced feedback (18.0\%) & Confirmation, expansion & Feedback that includes errors is associated with more retrospective reflection. Under unsuccessful feedback, participants' accounts agree with lower reflection-in-action. After successful overrides, deliberation concerns particular cases in which the AI may err. \\
\addlinespace
Behavior and outcome & Marginally lower adherence under successful feedback ($p = 0.064$), more adherence to incorrect AI advice under unsuccessful feedback ($p = 0.03$). No condition improves team performance (H\ref{hyp3} not supported) & Reduced reliance reported most often after successful overrides (11.5\%), increased reliance most often after unsuccessful overrides (16.4\%). Lessons about the task (50.3\%) far more frequent than a reliance strategy (8.2\%) & Confirmation, expansion & The reliance shifts appear in participants' own accounts in the same direction. Reflection mostly concerned participants' model of the task, which does not indicate which advice to override \\
\addlinespace
Perception & Change in perceived AI efficacy more favorable than in control in all feedback conditions, distinguishable from zero under successful and unsuccessful feedback and largest under unsuccessful feedback, where the mediated path to reliance is significant. Perceived self-efficacy shows no difference from control & More favorable view of the AI most frequent after unsuccessful overrides (14.8\%). Increased confidence most frequent after successful overrides (29.5\%), often phrased relative to the AI. Marginal influence of the feedback reported most often after unsuccessful overrides (39.3\%) & Confirmation for AI efficacy, discordance for self-efficacy and for the perceived influence of the feedback & Reliance under unsuccessful feedback is consistent with a revised belief about the AI. Confidence after successful overrides is relative to the AI and bound to cases, which the general scale does not ask about. The effects of unsuccessful feedback are only partly reflected in participants' own accounts \\
\bottomrule
\end{tabularx}
\end{table}

The two strands of evidence agree on reflection and on behavior and outcome, and they diverge on perception. The Discussion builds on these relations to explain how feedback on past overrides affects both forms of reflection (\Cref{sec:disc-reflection}), through which mechanisms it shifts reliance (\Cref{sec:discussion-reliance}), and why it does not improve performance (\Cref{sec:disc-task}).
\section{Discussion}
\label{sec-discussion}

Our work introduces feedback on past overrides as an intervention for reflection-on-action in human-AI collaboration and investigates its impact on reflection, reliance, and performance. Prior interventions such as explanations \citep{li2025text, solano2024explorable}, uncertainty displays \citep{holstein25balancing, zhang2020effect}, and cognitive forcing functions \citep{buccinca2021trust} have mainly focused on reflection-in-action (\Cref{tab:xai_interventions}). In addition, unique to our study is the distinction between balanced and unidirectional feedback. Unidirectional feedback bears high practical relevance, because in many domains feedback is selective by construction \citep{lakkaraju2017selective}, e.g. in credit lending repayment can only be observed for approved loans, so a review of a loan officer's overrides can only show the cases in which the officer approved a loan the AI advised rejecting. In this section, we discuss how feedback on past overrides affects reflection-in-action and reflection-on-action (\Cref{sec:disc-reflection}), through which mechanisms it shifts reliance behavior (\Cref{sec:discussion-reliance}), and why it does not improve performance (\Cref{sec:disc-task}). Each subsection relates the quantitative results to the reasoning behind our hypotheses and to participants' open-ended responses and then derives implications for research and design. Finally, we discuss the limitations of our study.

\subsection{Feedback on Past Overrides Raises Reflection-on-Action but Can Narrow Reflection-in-Action}
\label{sec:disc-reflection}

Our findings indicate that feedback on past overrides induces reflection-on-action (partially supporting H\ref{hyp1}), with participants in the balanced and unsuccessful feedback conditions reporting enhanced learning from errors. This pattern follows the reasoning behind H\ref{hyp1}, according to which feedback that reveals when and why the AI diverged from one's own decision provides the material for learning from errors \citep{hattie2007power, narciss2008feedback}. Feedback on successful overrides displays no errors of the participant, which is consistent with its smaller and non-significant increase. At the same time, unidirectional feedback is associated with lower reflection-in-action during subsequent decisions (significant for unsuccessful, marginal for successful feedback). Sensemaking theory provides a possible explanation for this pattern \citep{weick1995sensemaking}, which our hypotheses did not anticipate. Effective learning requires receiving information that challenges existing beliefs, creating psychological tension that ultimately results in adaptations of mental models that inform decision-making \citep{klein2006making, russell2024sensemaking, ehsan2024xai}. In this context, unidirectional explanatory feedback provides unambiguous signals about collaboration effectiveness that may have eliminated this cognitive dissonance. When participants consistently see evidence that they were wrong (unsuccessful feedback) or consistently right (successful feedback), compared to the AI, they may develop simplified collaboration heuristics instead of maintaining critical evaluation of each decision, reducing reflection-in-action during subsequent decisions. Balanced feedback, however, may maintain this productive cognitive dissonance required for sensemaking by presenting both successful and unsuccessful override outcomes. This balanced evidence, therefore, may preserve the psychological tension necessary for sustained reflection-in-action while still activating retrospective reflection-on-action processes.

The open-ended responses indicate that participants engaged less with feedback that showed only the AI's errors. Participants who received feedback on successful overrides least often referred to the individual cases shown to them, whereas participants with balanced feedback referred mainly to the cases they had gotten wrong. This difference fits the non-significant increase in reflection-on-action under successful feedback. Yet participants reflected on more than errors, describing improved understanding of the AI and the development of their decision strategy alongside a better understanding of the task. For reflection-in-action, the responses agree with the scale under feedback on unsuccessful overrides, where reports of a more deliberate approach are no more frequent than under balanced feedback. After successful overrides, participants report a more deliberate approach most often and describe weighing their own reasoning against the AI as often as analyzing cases more carefully. Feedback on successful overrides thus appears to direct deliberation toward particular cases in which the AI may err, which a scale of effortful processing across the whole task does not register as higher reflection-in-action.

These findings advance HCI research on human-AI collaboration and reflection, which has so far focused mainly on reflection-in-action \citep{buccinca2021trust, danry2023tell, gentile2025counterfactual, ma2025towards, zhang2020effect}. We show that feedback on past overrides can increase reflection-on-action and that a mechanism targeting one form of reflection also changes the other. \citet{schon1987educating} conceives of reflection-on-action as preparation for future reflection-in-action, whereas our findings suggest that this coupling can also turn negative, consistent with retrospective feedback replacing case-by-case deliberation with a simplified heuristic. Unidirectional feedback may thus remove the breakdowns that reflection requires \citep{baumer2015reflective, fleck2010reflecting, weick1995sensemaking}, whereas balanced feedback preserves them. Reflection interventions therefore need to be evaluated for both forms of reflection. For designers, this interaction matters because feedback in practice is often selective, for instance in credit lending, where repayment is only observed for approved loans and a review of a loan officer's overrides can therefore only cover overrides that led to an approval. However, system designers should be aware that feedback restricted in this way can raise reflection-on-action and at the same time lower reflection-in-action. Where the available outcomes allow it, a balanced review, showing an officer both the loans approved against the assessment that were repaid and those that defaulted, may avoid this narrowing.

\subsection{Unidirectional Feedback Shifts Reliance Through Different Mechanisms}
\label{sec:discussion-reliance}

Under the two unidirectional conditions, adherence shifts in opposite directions, as H\ref{hyp2} predicts. It is lower under feedback on successful overrides (marginally, $p = 0.064$) and descriptively higher under feedback on unsuccessful overrides ($p = 0.118$), and the two conditions differ from each other by a medium effect ($d = 0.59$).  H\ref{hyp2} rests on the reasoning that feedback on past overrides informs users about when to rely on their own judgment and when on the AI \citep{bailey2023meta, pescetelli2021role}. The mediation analyses support a belief update of this kind for feedback on unsuccessful overrides only, so the two shifts are symmetric in the behavioral data but asymmetric in their mechanism.

Feedback on unsuccessful overrides shifts adherence upward through what participants believe about the AI. The indirect effect through change in perceived AI efficacy explains the association in full and leaves nothing for a direct path, although adherence itself increases only descriptively and the total effect does not reach significance. This is the kind of competence calibration \citet{ma2024you} describe, but here it calibrates toward the wrong target, since the feedback reflects a biased sample of cases. By construction, participants see only those cases in which they overrode and the AI turned out to be right, so they update accurately on a sample that misrepresents the AI. Every case a participant saw showed the AI outperforming their own judgment, which leaves increased adherence as close to the only update the feedback permits. The condition therefore shows that participants update on the sample of cases they are shown, and not that they derive a reliance strategy through reflection. The consequence is a decision-maker who follows the AI more often, including when the AI is wrong, which is the pattern the context-dependent analysis isolates ($\beta = -0.83$, $p = 0.03$).

Feedback on successful overrides works differently, and its direct effect on reliance survives a belief-mediated path that runs the other way. Compared with the control condition, perceived AI efficacy declines less under feedback on successful overrides, descriptively also under balanced feedback, and increases under feedback on unsuccessful overrides, and participants with a more positive change in perceived AI efficacy adhere more to the AI advice. The mediated path therefore pushes adherence up while the direct path pushes it down, so the reduction in adherence occurs against the belief participants form about the AI and not through it. Perceived self-efficacy does not differ between conditions, which speaks against a general rise in confidence in one's own judgment as the driver of the behavior. What the direct path consists of remains open, because a null indirect effect speaks against the two global beliefs in our questionnaire but not against every belief a participant might revise. \citet{Bansal2019} describe users' mental models of the AI's error boundary, that is, beliefs about the kinds of cases in which the AI errs, and feedback that displays only errors of the AI may revise such a model locally.

The open-ended responses agree with these mechanisms, since participants' accounts of their reliance shift in the same directions as their behavior, with reduced reliance reported most often after successful overrides and increased reliance most often after unsuccessful overrides. When shown their own errors, participants revised their view of the AI more readily than their view of themselves. A more favorable view of the AI was reported most often after unsuccessful overrides, whereas reports of lower confidence in oneself were nearly absent, although this feedback displayed precisely the cases in which participants erred. This pattern fits the full mediation through perceived AI efficacy. After successful overrides, participants most often reported increased confidence in themselves. Many of them described it in relation to the AI and tied it to particular kinds of cases, such as applicants whose marital status the AI weighted too heavily. The general self-efficacy scale does not ask about such case-bound confidence, which may explain why it registers no change. These accounts point to a local revision of the AI's error boundary, although confirming it quantitatively requires case-level measures of perceived AI competence that our design does not provide.

Prior research on appropriate reliance traces miscalibrated reliance to two sources, the human, for example through self-confidence \citep{ma2024you, taudien2024metacognition} or an illusion of competence \citep{he2023knowing}, and the AI, for example through model updates that are incompatible with users' mental models \citep{bansal2019updates}. Our findings extend this view by pointing to feedback as a third source, since feedback on unsuccessful overrides miscalibrated reliance while the AI remained unchanged. Compatibility, which \citet{bansal2019updates} establish for model updates, is thus also a property of feedback. The findings also qualify explanations that attribute appropriate reliance to the match between humans' confidence in themselves and in the AI \citep{ma2024you, zahn2025calibration}, because reliance under feedback on successful overrides shifted against the measured belief about the AI. For explainable AI, they add a dimension to the design space, which has so far focused on the content \citep{danry2023tell, gentile2025counterfactual} and format \citep{cheng2019explaining, holstein25balancing} of explanations, namely the selection of which cases are fed back. With the explanation of each case held constant, different selections led to different effects on reflection and reliance. For designers, the selection of feedback cases is thus a decision with consequences for reliance. Feedback restricted to overrides keeps the review short but hands users a biased sample and omits the most frequent error, following incorrect AI advice. Feedback on unsuccessful adherence shows the AI failing where the user deferred to it and should lower adherence for the same reason that feedback on unsuccessful overrides raises it, so a review of every error of the human-AI team may correct this bias while covering only about a third of decisions.

\subsection{More Reflection Does Not Mean Better Reliance Decisions}
\label{sec:disc-task}

The reliance shifts do not translate into changes in human-AI team performance. H\ref{hyp3} assumed that feedback on past overrides improves performance by calibrating reliance, so that users follow the AI when its advice is correct and override it when it is wrong \citep{schemmer2023appropriate}. However, the context-dependent analysis (subset by AI advice correctness) points to a mechanism behind these patterns, where opposing effects among varying AI accuracies are buried on a global level. When AI advice is correct, participants receiving feedback about successful overrides decrease the human-AI team performance, marginally ($p = 0.095$). Conversely, when AI advice is incorrect, participants receiving feedback about unsuccessful overrides achieve lower human-AI team performance. Neither unidirectional condition thus shows the pattern H\ref{hyp3} presupposed, more adherence to correct and less adherence to incorrect advice. These opposing effects partially mask each other on a global level, which may explain the limited global performance improvements that led to the rejection of our third hypothesis (H\ref{hyp3}) on human-AI team performance. Balanced feedback shows neither error pattern and the highest adjusted performance, marginally above control ($p = 0.068$), which is the direction H\ref{hyp3} predicted for feedback that combines successful and unsuccessful overrides.

These findings on reliance patterns together with the dissonance of perceived increases in reflection-on-action and lowered reflection-in-action point to the risk of an \textit{illusion of learning} when unidirectional feedback is provided. While participants report higher reflection-on-action, they simultaneously report lower concurrent reflection-in-action, consistent with simplified reliance strategies, as adherence was descriptively higher under feedback on unsuccessful overrides and marginally lower under feedback on successful overrides. This is consistent with the context-dependent analyses, which show decreased performance. Our results point to this risk most clearly under feedback on unsuccessful overrides, where both shifts in reflection are significant. The illusion of learning extends prior literature that has highlighted the risk arising when perceived understanding does not align with objective understanding \citep{XUAN2025103376, cheng2019explaining} to feedback on past overrides. Our performance results suggest that activating reflection-on-action alone may be insufficient, since its effectiveness appears contingent on maintaining concurrent reflection-in-action.

The open-ended responses help explain why higher reflection-on-action did not lead to better reliance decisions. Participants most often reported lessons about the task, such as why they had misjudged a case or how much weight the applicants' attributes deserve, and few reported a strategy for when to follow or override the AI. A better model of the task does not indicate which advice to override, so reported reflection-on-action need not translate into better reliance decisions. Participants who received feedback on unsuccessful overrides also most often described the influence of the feedback as marginal, although this condition shows the clearest effects on reflection and on adherence to incorrect AI advice. Feedback can thus shape reliance in ways that users do not attribute to it.

Explainable AI research has studied explanations as support for evaluating a single piece of advice before a decision and has assessed them by their effect on that decision \citep{li2025text, schemmer2022meta_analysis_xai, solano2024explorable}. We extend this research by moving explanations into feedback on past overrides, after the outcome is known, where they no longer answer whether to follow a given piece of advice but show where the AI's errors lie relative to one's own. Unlike explanations in AI-provided feedback on task performance \citep{foerster2025feedback, spitzer2024transferring}, they are meant to support learning about the collaboration itself across cases \citep{Bansal2019, holstein2025development}. Our participants drew more lessons about the task than about the AI from them, which marks the gap between what such explanations offer and what users take from them. \citet{schoeffer2023interdependence, schoeffer2025ai} argue that reliance and accuracy are interdependent. We extend this argument to feedback interventions, whose opposing effects on cases with correct and incorrect AI advice can cancel out in the aggregate, which implies that intervention studies should report effects separately by the correctness of the AI advice. Our approach operationalizes the existing framework for analyzing reliance behavior of \citet{schoeffer2024explanations} by translating it into an actionable feedback intervention for participants.  Since participants in our study drew lessons mostly about the task, designers may need to make the reliance decision itself the object of such feedback, for example by summarizing in which kinds of cases the user's overrides succeeded and in which they failed. How feedback can support both forms of reflection without producing an illusion of learning is an open question for future research.

\subsection{Limitations and Future Research}
\label{sec:limitations}

\textbf{Reflection-in-action is measured retrospectively.} Participants reported on System 2 thinking after completing the task, so the measure records how they recall processing the decisions and not the processing itself. Both reflection measures were collected at the same point in time, so the temporal distinction between the two constructs is not mirrored in their measurement. Time spent on the decision pages and the pre-task cognitive reflection test both correlate with System 2 thinking and not with perceived learning from errors (\Cref{app:metric-validation}), which is the extent of the independent evidence available in this design. Participants' own reports and the measured effects also do not always align, since participants who received feedback on unsuccessful overrides most often reported that the feedback influenced them only marginally, although this condition shows the clearest effects on reflection-in-action and on adherence to incorrect AI advice. Process-tracing measures such as think-aloud protocols or eye-tracking capture concurrent processing directly, and a laboratory replication using them would test whether the reduction in reflection-in-action under unidirectional feedback appears outside self-report.

\textbf{Measurement considerations for reflection-on-action.} Our measure of retrospective reflection focused on error-based learning through items explicitly referencing mistakes (e.g., ``recognizing my errors,'' ``learning from errors''). That wording fits the conditions whose feedback displayed unsuccessful overrides, and it plausibly understates retrospective reflection in the successful condition, where the feedback displayed correct overrides. A participant who analyzed in depth why their own judgment outperformed the AI engaged in reflection-on-action and would still rate error-oriented items low, because the experience contained no recognizable errors. The intermediate adjusted mean of the successful condition ($\mu = 4.52$), between control ($\mu = 4.21$) and the two error-viewing conditions ($\mu = 4.68$), is consistent with this reading without establishing it. The open-ended responses support it. Participants who received feedback on successful overrides reported an improved understanding of the AI (29.5\%) and the development of a decision strategy (31.1\%) about as often as participants in the other feedback conditions, but an improved understanding of the task, which the error-oriented items address most directly, less often (36.1\% compared with 50.8\% and 63.9\%). A measure covering both valences of retrospective learning would separate a genuine gradient of engagement from an artifact of item wording.

\textbf{Feedback volume is endogenous to the behavior we study.} Feedback on overrides can only be delivered for cases in which a participant overrode the AI, so the number of feedback instances a participant received is a function of their own override rate, which is the very behavior the intervention is meant to shift. Including feedback exposure as a covariate does not fully resolve this, because that covariate is itself downstream of the disposition it is meant to control for. Since control participants receive no feedback by design, we examine feedback volume within the three feedback conditions. There, feedback count predicts none of the outcomes, and including it leaves the differences between feedback conditions in reliance unchanged (\Cref{app:feedback-frequency}). Participants who received more feedback did not report more learning from errors, which speaks against the amount of feedback as the driver of the reflection-on-action effect, although the design cannot vary the amount independently of participants' behavior. The dependency is structural. Any system that delivers feedback on overrides faces it, and a user who rarely overrides generates little material to reflect on. Future designs could enforce a minimum number of feedback cases or extend feedback to adherence cases.

\textbf{Our findings are bounded by the conditions of our study.}  Our study examined reflection-on-action under specific conditions, namely moderate AI accuracy (62.5\%), novice participants without domain expertise, brief collaboration periods of 18 decisions over approximately 15 minutes, and SHAP-based feature importance explanations. Domain expertise shapes how individuals process and integrate AI advice and explanations \citep{Zhao2025, morrison2024impact}, with experts drawing on prior knowledge and more developed mental models that enable more selective reliance than the novices in our study \citep{holstein2025development, Bansal2019}. AI accuracy shapes how trust forms and how it evolves during collaboration \citep{kahr2023seems}. Our moderate-performing system likely created different learning dynamics than participants would experience with either highly accurate or poor-performing AI. The explanations we provided further shaped this dynamic, as feature importance approaches like SHAP, which we presented as ranked feature lists, support particular cognitive processes that may interact differently with retrospective feedback on past overrides than alternative explanation types such as counterfactual explanations \citep{gentile2025counterfactual} or uncertainty and confidence information \citep{zhang2020effect, holstein25balancing}. Trust calibration and reliance strategies typically require extended interaction periods to fully develop and stabilize \citep{kahr2023seems}.

These contextual factors may help explain why reflection-on-action, while increased under balanced and unsuccessful feedback, did not translate into performance improvements. The temporal constraints of our design may have been insufficient for participants to develop and apply nuanced mental models based on retrospective feedback on past overrides. Our controlled experiment with a simplified loan approval task may not capture the complexity and stakes of real-world human-AI collaboration applications. Future research should therefore examine feedback on past overrides and reflection-on-action in field studies with domain experts using real-world AI systems, particularly investigating how long-term collaboration development unfolds in actual work environments where consequences extend beyond experimental participation. Longitudinal studies with repeated feedback over more rounds could also distinguish whether the effects of feedback on past overrides accumulate or reflect anchoring on the most recent feedback, which our three rounds could not separate.

\section{Conclusion}
\label{sec:conclusion}

This work introduces feedback on past overrides as a mechanism for reflection-on-action in human-AI collaboration, which research has so far addressed mainly through mechanisms for reflection-in-action such as explainable AI. Our findings show that both forms of reflection interact, since feedback that raises reflection after a decision can lower reflection during subsequent decisions. They also show that the selection of feedback cases matters, because feedback restricted to one type of override can itself miscalibrate reliance. Finally, more reflection does not by itself lead to better reliance decisions, since participants drew lessons mostly about the task and rarely about when to rely on the AI. Feedback that shows both successful and unsuccessful overrides avoided these pitfalls in our study, and designing feedback that makes the reliance decision itself the object of reflection remains a task for future research.

\section*{Generative AI Disclosure Use}

Generative AI tools were used at several stages of this work. ChatGPT (GPT-4.1), Claude Sonnet 4.5, and GitHub Copilot (with Claude Sonnet 4.5 and Gemini 3.1 Pro) were used to generate code for the visualizations. ChatGPT (GPT-4.1), DeepL Write, and Grammarly were used to improve the wording of the tutorials and explanations shown to participants. In the qualitative analysis, Claude (Opus 5.5) coded the open-ended responses outside the double-coded subset with the final codebook, and one author verified every code it assigned. Its agreement with both human coders is reported in \Cref{sec:experimental_design}. During the revision, Claude (Claude Sonnet 4.6 and 5.5 and Claude Opus 5.5), DeepL Write, and Grammarly were used to improve and refine the language of the manuscript and the response letter. The authors reviewed and verified all output of these tools and take full responsibility for the content of this article. 

\bibliographystyle{plainnat}
\bibliography{references}

\newpage
\appendix
\section*{Appendix}
\label{sec:Appendix}

\section{Statistical Reporting}

% LaTeX Table for Learning from Errors Results
\begin{table}[h!]
\centering
\caption{ANCOVA Results for Learning from Errors (Reflection-on-Action)}
\label{tab:learning_errors}
\begin{tabular}{lccccc}
\toprule
\textbf{Treatment} & \textbf{Adjusted Mean} & \textbf{SE} & \textbf{95\% CI} & \textbf{vs Control} & \textbf{Cohen's d} \\
& & & & \textbf{p-value}$^a$ & \\
\midrule
Control & 4.21 & 0.131 & [3.96, 4.47] & — & — \\
Balanced & 4.68 & 0.131 & [4.43, 4.94] & \textbf{0.018} & \textbf{0.46} \\
Successful & 4.52 & 0.131 & [4.27, 4.78] & 0.136 & 0.30 \\
Unsuccessful & 4.68 & 0.131 & [4.42, 4.94] & \textbf{0.018} & \textbf{0.46} \\
\midrule
\multicolumn{5}{l}{\textbf{Covariates}} & \textbf{p-value} \\
\midrule
\multicolumn{5}{l}{Cognitive Reflection} & 0.490 \\
\multicolumn{5}{l}{AI Efficacy} & \textbf{$<$ 0.001} \\
\multicolumn{5}{l}{Self-Efficacy} & 0.299 \\
\bottomrule
\end{tabular}
\begin{tablenotes}
\item \textit{Note}: N = 244 participants. Overall treatment effect: F(3,237) = 2.85, p = 0.038.
\item $^a$P-values are one-sided with Sidak adjustment, testing whether treatments increase learning from errors.
\item Significance codes: *** $p < 0.001$, ** $p < 0.01$, * $p < 0.05$
\end{tablenotes}
\end{table}

\begin{table}[h!]
\centering
\caption{ANCOVA Results for System 2 Thinking (Reflection-in-Action)}
\label{tab:system2_thinking}
\begin{tabular}{lccccc}
\toprule
\textbf{Treatment} & \textbf{Adjusted Mean} & \textbf{SE} & \textbf{95\% CI} & \textbf{vs Control} & \textbf{Cohen's d} \\
& & & & \textbf{p-value} & \\
\midrule
Control & 5.36 & 0.083 & [5.20, 5.52] & — & — \\
Balanced & 5.32 & 0.083 & [5.16, 5.48] & 0.984 & 0.06 \\
Successful & 5.10 & 0.083 & [4.93, 5.26] & 0.077 & 0.41 \\
Unsuccessful & 5.06 & 0.083 & [4.90, 5.22] & \textbf{0.034} & \textbf{0.46} \\
\midrule
\multicolumn{5}{l}{\textbf{Covariates}} & \textbf{p-value} \\
\midrule
\multicolumn{5}{l}{Cognitive Reflection} & \textbf{0.015} \\
\multicolumn{5}{l}{AI Efficacy} & 0.079 \\
\multicolumn{5}{l}{Self-Efficacy} & \textbf{0.019} \\
\bottomrule
\end{tabular}
\begin{tablenotes}
\item \textit{Note}: N = 244 participants. F(3,237) = 3.39, p = 0.019. Covariates: cognitive reflection degree,  AI efficacy,  self-efficacy. P-values are two-sided with Sidak adjustment for multiple comparisons.
\end{tablenotes}
\Description{This table presents ANCOVA results for System 2 thinking, specifically reflection-in-action, comparing different treatment conditions against a control group. The table contains six columns showing treatment condition, adjusted mean, standard error, 95\% confidence interval, p-value compared to control, and Cohen's d effect size. The control group shows an adjusted mean of 5.36 with standard error 0.083 and 95\% confidence interval from 5.20 to 5.52, with no comparison values since it serves as the reference. The balanced treatment condition has an adjusted mean of 5.32, standard error 0.083, 95\% confidence interval from 5.16 to 5.48, p-value of 0.984 compared to control, and Cohen's d of 0.06 indicating a negligible effect size. The successful treatment condition shows adjusted mean 5.10, standard error 0.083, 95\% confidence interval from 4.93 to 5.26, p-value of 0.077 compared to control, and Cohen's d of 0.41 indicating a small effect size. The unsuccessful treatment condition has adjusted mean 5.06, standard error 0.083, 95\% confidence interval from 4.90 to 5.22, p-value of 0.034 compared to control indicating statistical significance, and Cohen's d of 0.46 indicating a small effect size. The analysis included 244 participants with overall model significance of F 3 comma 237 equals 3.39 and p-value 0.019. The model included covariates for cognitive reflection degree, AI efficacy, and self-efficacy. The p-values reported are two-sided with Sidak adjustment for multiple comparisons.}
\end{table}

\begin{table}[h!]
\centering
\caption{ANCOVA Results for AI Reliance (Round 3)}
\label{tab:reliance_round3}
\begin{tabular}{lccccc}
\toprule
\textbf{Treatment} & \textbf{Adjusted Mean} & \textbf{SE} & \textbf{95\% CI} & \textbf{vs Control} & \textbf{Cohen's d} \\
& & & & \textbf{p-value}$^a$ & \\
\midrule
Control & 0.724 & 0.035 & [0.656, 0.793] & — & — \\
Balanced & 0.764 & 0.033 & [0.699, 0.829] & 0.480 & 0.20 \\
Successful & 0.656 & 0.026 & [0.604, 0.708] & 0.064 & 0.33 \\
Unsuccessful & 0.776 & 0.026 & [0.725, 0.828] & 0.118 & 0.26 \\
\midrule
\multicolumn{5}{l}{Cognitive Reflection} & 0.364 \\
\multicolumn{5}{l}{AI Efficacy} & 0.461 \\
\multicolumn{5}{l}{Self-Efficacy} & 0.100 \\
\multicolumn{5}{l}{Baseline Reliance (Round 1)} & 0.142 \\
\multicolumn{5}{l}{Feedback Exposure (after Round 1)} & 0.665 \\
\multicolumn{5}{l}{Feedback Exposure (after Round 2)} & 0.507 \\
\bottomrule
\end{tabular}
\begin{tablenotes}
\item \textit{Note}: N = 244 participants. F(3,234) = 4.09, p = 0.007. Covariates: cognitive reflection degree,  AI efficacy,  self-efficacy, baseline reliance (Round 1), feedback exposure (Rounds 1 \& 2). 
\item $^a$P-values: one-sided for Successful and Unsuccessful conditions (testing directional hypotheses), two-sided for Balanced condition.
\end{tablenotes}
\end{table}

\begin{table}[h!]
\centering
\caption{ANCOVA Results for Human-AI Team Performance (Round 3)}
\label{tab:performance_round3}
\begin{tabular}{lccccc}
\toprule
\textbf{Treatment} & \textbf{Adjusted Mean} & \textbf{SE} & \textbf{95\% CI} & \textbf{vs Control} & \textbf{Cohen's d} \\
& & & & \textbf{p-value}$^a$ & \\
\midrule
Control & 0.613 & 0.034 & [0.546, 0.679] & — & — \\
Balanced & 0.692 & 0.032 & [0.629, 0.756] & 0.068 & 0.37 \\
Successful & 0.612 & 0.028 & [0.557, 0.667] & 0.506 & 0.00 \\
Unsuccessful & 0.588 & 0.028 & [0.533, 0.643] & 0.713 & 0.11 \\
\midrule
\multicolumn{5}{l}{Cognitive Reflection} & 0.769 \\
\multicolumn{5}{l}{AI Efficacy} & 0.411 \\
\multicolumn{5}{l}{Self-Efficacy} & 0.779 \\
\multicolumn{5}{l}{Baseline Performance (Round 1)} & 0.877 \\
\multicolumn{5}{l}{Feedback Exposure (after Round 1)} & 0.488 \\
\multicolumn{5}{l}{Feedback Exposure (after Round 2)} & 0.110 \\
\bottomrule
\end{tabular}
\begin{tablenotes}
\item \textit{Note}: N = 244 participants. F(3,234) = 2.15, p = 0.095. Covariates: cognitive reflection degree,  AI efficacy,  self-efficacy, baseline performance (Round 1), feedback exposure (Rounds 1 \& 2). 
\item $^a$P-values are one-sided, testing whether treatments increase performance compared to control.
\end{tablenotes}
\end{table}

\clearpage
\section{Validation of Reflection Measures}
\label{app:metric-validation}

Our two reflection measures---perceived learning from errors (reflection-on-action) and System 2 thinking (reflection-in-action)---are both self-reported in the post-task questionnaire. This appendix reports the analyses establishing that they are empirically distinct and that each is anchored by evidence collected independently of self-report.

\textbf{Distinctness of the two measures.} \Cref{tab:construct-validity} reports the correlation between the two measures within each condition. In the control condition, which receives no feedback and is therefore free of treatment effects, the association is small and not significant. The correlation is substantially larger in the unsuccessful condition, where the feedback made errors salient and cognitive effort and perceived learning from errors became coupled. This is a treatment-induced association rather than evidence of construct overlap at baseline.

\begin{table}[h!]
\centering
\caption{Correlation between System 2 thinking (reflection-in-action) and perceived learning from errors (reflection-on-action), by condition.}
\label{tab:construct-validity}
\begin{tabular}{lccc}
\toprule
\textbf{Condition} & \textbf{$r$} & \textbf{95\% CI} & \textbf{$p$} \\
\midrule
Control       & 0.171 & [-0.085, 0.405] & 0.189 \\
Balanced      & 0.255 & [0.003, 0.476]  & 0.047 \\
Successful    & 0.267 & [0.016, 0.486]  & 0.037 \\
Unsuccessful  & 0.602 & [0.413, 0.741]  & $<$ 0.001 \\
\midrule
Pooled        & 0.292 & [0.172, 0.403]  & $<$ 0.001 \\
\bottomrule
\end{tabular}
\Description{Table reporting Pearson correlations between System 2 thinking and perceived learning from errors within each of the four experimental conditions and pooled across conditions, with 95 percent confidence intervals and p-values. The control condition shows r equals 0.171 with a confidence interval from negative 0.085 to 0.405 and p equals 0.189. The balanced condition shows r equals 0.255 with a confidence interval from 0.003 to 0.476 and p equals 0.047. The successful condition shows r equals 0.267 with a confidence interval from 0.016 to 0.486 and p equals 0.037. The unsuccessful condition shows r equals 0.602 with a confidence interval from 0.413 to 0.741 and p below 0.001. Pooled across conditions, r equals 0.292 with a confidence interval from 0.172 to 0.403 and p below 0.001.}
\end{table}

\textbf{Objective anchors.} \Cref{tab:objective-anchors} relates both self-reported measures to two objective quantities. The first is the cognitive reflection test \citep{frederick2005cognitive}, collected before any task exposure and balanced across conditions ($p = 0.658$). The second is the time participants spent on the decision pages, derived from page-level timestamps recorded by the survey platform. Timing analyses use all participants, and time on the feedback pages is available only for the 183 participants in the feedback conditions. Both objective measures predict System 2 thinking and are unrelated to perceived learning from errors, which is what the temporal separation of the two forms of reflection implies. Time on the feedback pages shows the reverse tendency as it does not correlate with System 2 thinking and shows a small association with perceived learning from errors that does not reach significance ($r = 0.142$, $p = 0.055$).

\begin{table}[h!]
\centering
\caption{Self-reported reflection measures against objective anchors. Time on the feedback pages is recorded only in the three feedback conditions.}
\label{tab:objective-anchors}
\begin{tabular}{llrcrr}
\toprule
\textbf{Objective measure} & \textbf{Self-reported measure} & \textbf{$r$} & \textbf{95\% CI} & \textbf{$p$} & \textbf{$n$} \\
\midrule
Cognitive reflection test & System 2 thinking    & \phantom{-}0.147 & [\phantom{-}0.022, \phantom{-}0.268] & \textbf{0.022} & 244 \\
Cognitive reflection test & System 1 thinking    & -0.139 & [-0.260, -0.014] & \textbf{0.030} & 244 \\
Cognitive reflection test & Learning from errors & \phantom{-}0.045 & [-0.081, \phantom{-}0.169] & 0.487 & 244 \\
\midrule
Time on decision pages    & System 2 thinking    & \phantom{-}0.198 & [\phantom{-}0.074, \phantom{-}0.315] & \textbf{0.002} & 244 \\
Time on decision pages    & Learning from errors & \phantom{-}0.096 & [-0.030, \phantom{-}0.219] & 0.134 & 244 \\
\midrule
Time on feedback pages    & Learning from errors & \phantom{-}0.142 & [-0.003, \phantom{-}0.282] & 0.055 & 183 \\
Time on feedback pages    & System 2 thinking    & \phantom{-}0.074 & [-0.072, \phantom{-}0.217] & 0.317 & 183 \\
\bottomrule
\end{tabular}
\Description{Table relating the self-reported reflection measures to objective anchors. The cognitive reflection test correlates with System 2 thinking at r equals 0.147 with a confidence interval from 0.022 to 0.268, p equals 0.022, n equals 244. It correlates negatively with System 1 thinking at r equals negative 0.139 with a confidence interval from negative 0.260 to negative 0.014, p equals 0.030, n equals 244. It shows no relationship with learning from errors at r equals 0.045 with a confidence interval from negative 0.081 to 0.169, p equals 0.487, n equals 244. Time spent on decision pages correlates with System 2 thinking at r equals 0.198 with a confidence interval from 0.074 to 0.315, p equals 0.002, n equals 244, and shows no relationship with learning from errors at r equals 0.096 with a confidence interval from negative 0.030 to 0.219, p equals 0.134, n equals 244. Time spent on feedback pages, recorded only for the 183 participants in the three feedback conditions, shows a small association with learning from errors at r equals 0.142 with a confidence interval from negative 0.003 to 0.282 and p equals 0.055, and no relationship with System 2 thinking at r equals 0.074 with a confidence interval from negative 0.072 to 0.217 and p equals 0.317.}
\end{table}

\textbf{System 1 thinking.} Heuristic System 1 thinking does not differ across conditions ($F(3,240) = 0.118$, $p = 0.949$). It behaves as a stable disposition where participants scoring higher on System 1 thinking report higher post-task AI efficacy ($r = 0.167$, $p = 0.009$), while it is unrelated to the time spent on decision pages ($r = -0.051$, $p = 0.424$).

\clearpage
\section{Longitudinal Analysis Across Rounds}
\label{app:longitudinal}

The main text analyzes round 3, after full feedback exposure. This appendix reports the round 2 results and a model that tests whether the treatment effect differs between rounds.

\textbf{Round 2.} Replicating the round 3 ANCOVA specification on round 2 reliance, with the number of feedback instances shown after round 1 as the feedback covariate, yields no significant main effect of treatment condition ($F(3,235) = 1.66$, $p = 0.176$), and feedback count does not predict reliance ($p = 0.812$). In the planned contrasts, participants receiving feedback on successful overrides already adhere less to AI advice than control participants ($\Delta = -0.078$, $p = 0.021$, one-sided, $d = 0.42$), while the unsuccessful ($p = 0.832$, one-sided) and balanced ($p = 0.527$, two-sided) conditions do not differ from control. The reduction in adherence under successful feedback thus appears after the first wave of feedback, whereas the increase under unsuccessful feedback emerges only by round 3, in line with the unsuccessful $\times$ round term in the model below.

\textbf{Treatment $\times$ Round.} We estimate a mixed-effects logistic regression on individual decisions in rounds 2 and 3, with treatment condition, round, and their interaction as fixed effects, round 1 reliance, cognitive reflection degree, and pre-task AI and self-efficacy as covariates, and random intercepts for participants (\Cref{tab:glmm-rounds}). Control and round 2 serve as reference levels. The Treatment $\times$ Round interaction is not significant ($\chi^2(3) = 5.45$, $p = 0.142$), and neither is the treatment term at the round 2 reference level ($\chi^2(3) = 4.10$, $p = 0.251$). Among the individual interaction terms, only the unsuccessful condition approaches significance ($b = 0.459$, $SE = 0.243$, $p = 0.058$), in the direction of an effect that grows from round 2 to round 3. The model neither establishes that the treatment effect is constant across rounds nor that it changes.

\textbf{Unadjusted model.} A simpler specification without covariates, estimated on all three rounds, reaches the same conclusion about the interaction ($\chi^2(6) = 9.88$, $p = 0.130$). We report no treatment main effect from either interaction model. Under treatment contrasts, a Type III main effect is evaluated at the reference level of the other factor, so in the three-round model that term contrasts conditions in round 1, before any feedback was delivered. Reliance in round 1 does differ across conditions ($F(3,240) = 3.09$, $p = 0.028$), and every model in this paper adjusts for it.

\begin{table}[h!]
\centering
\small
\caption{Covariate-adjusted mixed-effects logistic regression on reliance, rounds 2 and 3. Reference levels are the control condition and round 2.}
\label{tab:glmm-rounds}
\begin{tabular}{@{}lrrrr@{}}
\toprule
\textbf{Fixed effect} & \textbf{Estimate} & \textbf{SE} & \textbf{$z$} & \textbf{$p$} \\
\midrule
(Intercept)                        & \phantom{-}0.740 & 0.284 & \phantom{-}2.61 & \textbf{0.009} \\
Balanced                           & -0.094 & 0.183 & -0.52 & 0.607 \\
Successful                         & -0.344 & 0.178 & -1.94 & 0.053 \\
Unsuccessful                       & -0.173 & 0.181 & -0.96 & 0.339 \\
Round 3                            & -0.130 & 0.170 & -0.76 & 0.445 \\
Baseline reliance (round 1)        & \phantom{-}0.455 & 0.266 & \phantom{-}1.71 & 0.088 \\
Cognitive reflection degree        & \phantom{-}0.062 & 0.123 & \phantom{-}0.51 & 0.611 \\
AI efficacy (pre-task)             & \phantom{-}0.186 & 0.262 & \phantom{-}0.71 & 0.479 \\
Self-efficacy (pre-task)           & -0.163 & 0.222 & -0.73 & 0.463 \\
Balanced $\times$ Round 3          & \phantom{-}0.279 & 0.242 & \phantom{-}1.15 & 0.249 \\
Successful $\times$ Round 3        & -0.009 & 0.232 & -0.04 & 0.968 \\
Unsuccessful $\times$ Round 3      & \phantom{-}0.459 & 0.243 & \phantom{-}1.89 & 0.058 \\
\bottomrule
\end{tabular}
\begin{tablenotes}
\item \textit{Note}: N = 2{,}928 observations from 244 participants. Random intercept for participant, $\sigma^2 = 0.111$ ($SD = 0.333$). Type III tests: Treatment $\chi^2(3) = 4.10$, $p = 0.251$; Round $\chi^2(1) = 0.58$, $p = 0.445$; Treatment $\times$ Round $\chi^2(3) = 5.45$, $p = 0.142$.
\end{tablenotes}
\end{table}

\clearpage
\section{Robustness to Feedback Frequency}
\label{app:feedback-frequency}

Feedback is generated from participants' own overrides, so the number of feedback instances each participant received is determined by their behavior and not by the design. Feedback volume differs across conditions in both waves ($p < 0.001$). Participants in the control condition receive no feedback by design, so their feedback count is zero, and a model that includes the count across all four conditions largely re-codes the contrast with control instead of separating the kind of feedback from its amount \citep{rosenbaum1984consequences, montgomery2018conditioning}. We therefore assess the role of feedback volume within the three feedback conditions ($n = 183$) and re-estimate the round 3 ANCOVA for each outcome with and without the exact number of feedback instances a participant received as a covariate (\Cref{tab:feedback-robustness}). Both specifications include cognitive reflection degree, pre-task AI and self-efficacy, and round 1 reliance (round 1 performance for the performance outcome), so the columns are comparable with each other but not with the models in \Cref{sec-results}, which include the control condition.

\begin{table}[h!]
\centering
\small
\caption{Differences between the three feedback conditions in the round 3 ANCOVA with and without the number of feedback instances as a covariate ($n = 183$; control excluded).}
\label{tab:feedback-robustness}
\begin{tabular}{@{}lccr@{}}
\toprule
\textbf{Outcome} & \textbf{Without feedback count} & \textbf{With feedback count} & \textbf{Feedback count term} \\
\midrule
Reliance              & $F(2,176) = 6.26$, $p = 0.002$ & $F(2,175) = 6.30$, $p = 0.002$ & $b = -0.007$ (0.014), $p = 0.629$ \\
Reflection-in-action  & $F(2,176) = 2.49$, $p = 0.086$ & $F(2,175) = 1.84$, $p = 0.162$ & $b = -0.000$ (0.009), $p = 0.962$ \\
Reflection-on-action  & $F(2,176) = 0.53$, $p = 0.591$ & $F(2,175) = 0.45$, $p = 0.641$ & $b = \phantom{-}0.016$ (0.013), $p = 0.236$ \\
Performance           & $F(2,176) = 1.79$, $p = 0.169$ & $F(2,175) = 2.84$, $p = 0.061$ & $b = -0.020$ (0.013), $p = 0.113$ \\
\bottomrule
\end{tabular}
\Description{Table comparing the differences between the three feedback conditions, with the control condition excluded and 183 participants, in the round three ANCOVA with and without the number of feedback instances as a covariate. For reliance, the effect is F of 2 and 176 equals 6.26 with p equals 0.002 without the covariate and F of 2 and 175 equals 6.30 with p equals 0.002 with it, and the feedback count coefficient is negative 0.007 with standard error 0.014 and p equals 0.629. For reflection-in-action, the effect is F of 2 and 176 equals 2.49 with p equals 0.086 without the covariate and F of 2 and 175 equals 1.84 with p equals 0.162 with it, and the feedback count coefficient is 0.000 with standard error 0.009 and p equals 0.962. For reflection-on-action, the effect is F of 2 and 176 equals 0.53 with p equals 0.591 without the covariate and F of 2 and 175 equals 0.45 with p equals 0.641 with it, and the feedback count coefficient is 0.016 with standard error 0.013 and p equals 0.236. For performance, the effect is F of 2 and 176 equals 1.79 with p equals 0.169 without the covariate and F of 2 and 175 equals 2.84 with p equals 0.061 with it, and the feedback count coefficient is negative 0.020 with standard error 0.013 and p equals 0.113.}
\end{table}

Feedback count predicts none of the four outcomes (all $p > 0.11$). The differences between feedback conditions in reliance are unchanged when the count is included. For reflection-on-action, the three feedback conditions do not differ from each other with or without the count, since the effect reported in \Cref{sec:cognitive_processes} lies in the contrast with control. Participants who received more feedback did not report more learning from errors than participants who received less, so the reflection-on-action effect reflects receiving feedback on past overrides and not the amount of it. For reflection-in-action, the difference between feedback conditions is small in this subset and weakens further with the count. We discuss the dependency of feedback volume on behavior in \Cref{sec:limitations}.

\textbf{Symmetry within the balanced condition.} The balanced condition is balanced by design and not by realized exposure, because which cases a participant sees depends on the outcomes of their own overrides. Participants in this condition saw on average $1.64$ ($SD = 1.02$) successful and $1.28$ ($SD = 1.31$) unsuccessful override cases across the two feedback waves, a descriptive skew toward successful overrides that does not reach significance ($t(60) = 1.65$, $p = 0.103$).

\clearpage
\section{Mediation Analyses}
\label{app:mediation}

This appendix reports the mediator and outcome models underlying the causal mediation results in \Cref{tab:mediation}. For each participant we compute delta scores as post-task minus pre-task values of perceived AI efficacy and perceived self-efficacy. Each contrast compares one feedback condition against control ($n = 122$). The mediator model regresses the delta score on the treatment indicator, giving path $a$. The outcome model regresses reliance in round 3 on the treatment indicator and the delta score, giving the direct path $c'$ and the mediator path $b$.

\begin{table}[h!]
\centering
\small
\caption{Causal mediation of reliance in round 3, each feedback condition against control, with 1,000 quasi-Bayesian simulations. ACME denotes the average causal mediation effect and ADE the average direct effect.}
\label{tab:mediation}
\begin{tabular}{@{}llrcr@{}}
\toprule
\textbf{Condition} & \textbf{Effect} & \textbf{Estimate} & \textbf{95\% CI} & \textbf{$p$} \\
\midrule
\multicolumn{5}{@{}l}{\textit{Mediator: change in perceived AI efficacy}} \\
\addlinespace[2pt]
Balanced     & ACME  & 0.017  & [-0.006, \phantom{-}0.046] & 0.160 \\
             & ADE   & 0.021  & [-0.050, \phantom{-}0.096] & 0.554 \\
             & Total & 0.039  & [-0.032, \phantom{-}0.115] & 0.324 \\
\addlinespace[2pt]
Successful   & ACME  & 0.024  & [\phantom{-}0.001, \phantom{-}0.053] & \textbf{0.038} \\
             & ADE   & -0.097 & [-0.168, -0.021] & \textbf{0.006} \\
             & Total & -0.073 & [-0.144, \phantom{-}0.000] & 0.052 \\
\addlinespace[2pt]
Unsuccessful & ACME  & 0.040  & [\phantom{-}0.013, \phantom{-}0.076] & \textbf{$<$ 0.001} \\
             & ADE   & 0.016  & [-0.048, \phantom{-}0.086] & 0.648 \\
             & Total & 0.056  & [-0.012, \phantom{-}0.126] & 0.108 \\
\midrule
\multicolumn{5}{@{}l}{\textit{Mediator: change in perceived self-efficacy}} \\
\addlinespace[2pt]
Balanced     & ACME  & 0.001  & [-0.011, \phantom{-}0.016] & 0.812 \\
             & ADE   & 0.038  & [-0.031, \phantom{-}0.112] & 0.318 \\
             & Total & 0.039  & [-0.031, \phantom{-}0.115] & 0.300 \\
\addlinespace[2pt]
Successful   & ACME  & -0.002 & [-0.016, \phantom{-}0.007] & 0.716 \\
             & ADE   & -0.073 & [-0.146, -0.003] & \textbf{0.044} \\
             & Total & -0.075 & [-0.150, -0.004] & \textbf{0.036} \\
\addlinespace[2pt]
Unsuccessful & ACME  & 0.003  & [-0.005, \phantom{-}0.017] & 0.542 \\
             & ADE   & 0.054  & [-0.016, \phantom{-}0.124] & 0.128 \\
             & Total & 0.057  & [-0.014, \phantom{-}0.130] & 0.116 \\
\bottomrule
\end{tabular}
\end{table}

\begin{table}[h!]
\centering
\small
\caption{Mediator and outcome model coefficients underlying the mediation analyses. Standard errors in parentheses.}
\label{tab:mediation-paths}
\begin{tabular}{@{}lrrr@{}}
\toprule
\textbf{Condition vs.\ control} & \textbf{$a$ (T $\rightarrow$ M)} & \textbf{$b$ (M $\rightarrow$ Y)} & \textbf{$c'$ (direct)} \\
\midrule
\multicolumn{4}{@{}l}{\textit{Mediator: change in perceived AI efficacy}} \\
\addlinespace[2pt]
Balanced     & \phantom{-}0.046 (0.032) & \phantom{-}0.389 (0.102) & \phantom{-}0.023 (0.036) \\
Successful   & \phantom{-}0.058 (0.028) & \phantom{-}0.413 (0.117) & -0.098 (0.036) \\
Unsuccessful & \phantom{-}0.127 (0.031) & \phantom{-}0.315 (0.099) & \phantom{-}0.017 (0.036) \\
\midrule
\multicolumn{4}{@{}l}{\textit{Mediator: change in perceived self-efficacy}} \\
\addlinespace[2pt]
Balanced     & -0.008 (0.029) & -0.173 (0.117) & \phantom{-}0.040 (0.037) \\
Successful   & -0.020 (0.033) & \phantom{-}0.118 (0.104) & -0.071 (0.037) \\
Unsuccessful & -0.031 (0.036) & -0.099 (0.089) & \phantom{-}0.054 (0.035) \\
\bottomrule
\end{tabular}
\Description{Table reporting the mediator and outcome model coefficients underlying the mediation analyses, with standard errors in parentheses. With change in perceived AI efficacy as the mediator, the balanced condition has an a path of 0.046 with standard error 0.032, a b path of 0.389 with standard error 0.102, and a direct path of 0.023 with standard error 0.036. The successful condition has an a path of 0.058 with standard error 0.028, a b path of 0.413 with standard error 0.117, and a direct path of negative 0.098 with standard error 0.036. The unsuccessful condition has an a path of 0.127 with standard error 0.031, a b path of 0.315 with standard error 0.099, and a direct path of 0.017 with standard error 0.036. With change in perceived self-efficacy as the mediator, the balanced condition has an a path of negative 0.008, a b path of negative 0.173, and a direct path of 0.040. The successful condition has an a path of negative 0.020, a b path of 0.118, and a direct path of negative 0.071. The unsuccessful condition has an a path of negative 0.031, a b path of negative 0.099, and a direct path of 0.054.}
\end{table}

The $a$ paths are positive in all three feedback conditions, so perceived AI efficacy develops more favorably than in the control condition. The shift is largest under unsuccessful feedback, exceeds twice its standard error under successful feedback, and is not distinguishable from zero under balanced feedback ($a = 0.046$, $SE = 0.032$). The $b$ paths are positive and of comparable magnitude across conditions, so participants whose perception of the AI improves rely on it more. The combination of a positive $a$ path and a negative $c'$ path in the successful condition produces the competitive mediation reported in \Cref{sec:mediation-results}. For perceived self-efficacy, the $a$ paths are small and not distinguishable from zero, which is why no mediated effect through this construct reaches significance.

\clearpage
\section{Qualitative Reporting}
\label{app:qual}

\begin{table}[!h]
\centering
\footnotesize
\renewcommand{\arraystretch}{1}
\setlength{\tabcolsep}{5pt}
\caption{Prevalence of themes and codes across feedback conditions. Cells report the percentage of participants
whose open-ended responses carried the theme or code at least once. The table also documents the codebook, listing each facet with its themes (in bold) and codes. The nine task-detail codes are reported as a single aggregate.}
\label{tab:prevalence-full}
\begin{tabular}{@{}lrrrr@{}}
\toprule
Theme / code & Balanced & Successful & Unsuccessful & All \\
 & ($n=61$) & ($n=61$) & ($n=61$) & ($n=183$) \\
\midrule
\multicolumn{5}{@{}l}{\textit{Facet 1: Decision cues}} \\
\addlinespace[1pt]
\textbf{Feedback: Outcomes} & \textbf{45.9} & \textbf{49.2} & \textbf{47.5} & \textbf{47.5} \\
\quad General override correctness & 31.1 & 36.1 & 31.1 & 32.8 \\
\quad Actual repayment outcome & 1.6 & 6.6 & 11.5 & 6.6 \\
\quad Number of overrides shown & 26.2 & 24.6 & 31.1 & 27.3 \\
\addlinespace[1pt]
\textbf{Feedback: Instances} & \textbf{47.5} & \textbf{42.6} & \textbf{55.7} & \textbf{48.6} \\
\quad All feedback instances & 26.2 & 0.0 & 0.0 & 8.7 \\
\quad Feedback instances with own errors & 23.0 & 0.0 & 55.7 & 26.2 \\
\quad Feedback instances with AI errors & 4.9 & 42.6 & 0.0 & 15.8 \\
\addlinespace[1pt]
\textbf{AI Output} & \textbf{31.1} & \textbf{36.1} & \textbf{29.5} & \textbf{32.2} \\
\quad AI recommendation & 11.5 & 23.0 & 8.2 & 14.2 \\
\quad AI explanation & 23.0 & 19.7 & 23.0 & 21.9 \\
\addlinespace[1pt]
\textbf{Task details} & \textbf{50.8} & \textbf{63.9} & \textbf{65.6} & \textbf{60.1} \\
\quad Task details & 50.8 & 63.9 & 65.6 & 60.1 \\
\addlinespace[1pt]
\midrule\midrule
\multicolumn{5}{@{}l}{\textit{Facet 2: Effects}} \\
\addlinespace[1pt]
\textbf{Reliance behavior changed} & \textbf{13.1} & \textbf{13.1} & \textbf{16.4} & \textbf{14.2} \\
\quad Reliance on AI increased & 8.2 & 1.6 & 16.4 & 8.7 \\
\quad Reliance on AI decreased & 4.9 & 11.5 & 0.0 & 5.5 \\
\addlinespace[1pt]
\textbf{Perceptions of AI and self changed} & \textbf{29.5} & \textbf{41.0} & \textbf{29.5} & \textbf{33.3} \\
\quad AI efficacy increased & 6.6 & 1.6 & 14.8 & 7.7 \\
\quad AI efficacy decreased & 4.9 & 13.1 & 3.3 & 7.1 \\
\quad AI efficacy adjusted & 6.6 & 6.6 & 1.6 & 4.9 \\
\quad Self-efficacy increased & 14.8 & 29.5 & 9.8 & 18.0 \\
\quad Self-efficacy decreased & 1.6 & 0.0 & 3.3 & 1.6 \\
\quad Self-efficacy adjusted & 3.3 & 1.6 & 0.0 & 1.6 \\
\addlinespace[1pt]
\textbf{Reflection on Action: Task understanding improved} & \textbf{50.8} & \textbf{36.1} & \textbf{63.9} & \textbf{50.3} \\
\quad Own task error reflected on & 16.4 & 6.6 & 34.4 & 19.1 \\
\quad Attribute weighting reconsidered & 13.1 & 14.8 & 24.6 & 17.5 \\
\quad General task knowledge gained & 31.1 & 21.3 & 23.0 & 25.1 \\
\addlinespace[1pt]
\textbf{Reflection on Action: AI understanding improved} & \textbf{26.2} & \textbf{29.5} & \textbf{26.2} & \textbf{27.3} \\
\quad Own reliance behavior reflected on & 13.1 & 11.5 & 6.6 & 10.4 \\
\quad AI decision logic reflected on & 8.2 & 11.5 & 11.5 & 10.4 \\
\quad Logic behind single AI decision understood & 6.6 & 6.6 & 4.9 & 6.0 \\
\quad General AI decision logic inferred from observation & 6.6 & 3.3 & 6.6 & 5.5 \\
\addlinespace[1pt]
\textbf{Reflection on Action: Strategy developed} & \textbf{29.5} & \textbf{31.1} & \textbf{36.1} & \textbf{32.2} \\
\quad Task completion logic simplified & 0.0 & 0.0 & 1.6 & 0.5 \\
\quad Task completion logic advanced & 4.9 & 3.3 & 8.2 & 5.5 \\
\quad Reliance strategy developed & 6.6 & 9.8 & 8.2 & 8.2 \\
\quad Existing strategy reinforced & 14.8 & 19.7 & 6.6 & 13.7 \\
\quad Strategy changed unspecified & 4.9 & 4.9 & 11.5 & 7.1 \\
\addlinespace[1pt]
\textbf{Reflection in Action: Task approached more deliberately} & \textbf{18.0} & \textbf{26.2} & \textbf{16.4} & \textbf{20.2} \\
\quad Own reasoning weighed against AI & 8.2 & 14.8 & 3.3 & 8.7 \\
\quad Cases analyzed more carefully & 13.1 & 14.8 & 13.1 & 13.7 \\
\addlinespace[1pt]
\textbf{Marginally influenced} & \textbf{24.6} & \textbf{31.1} & \textbf{39.3} & \textbf{31.7} \\
\quad Own judgment remained decisive over AI & 8.2 & 13.1 & 9.8 & 10.4 \\
\quad No or little influence reported & 23.0 & 23.0 & 34.4 & 26.8 \\
\addlinespace[1pt]
\textbf{Feedback expected to be more elaborate} & \textbf{4.9} & \textbf{3.3} & \textbf{9.8} & \textbf{6.0} \\
\quad Feedback not understood & 1.6 & 0.0 & 4.9 & 2.2 \\
\quad Explanation lacked detail & 4.9 & 3.3 & 4.9 & 4.4 \\
\addlinespace[1pt]
\bottomrule
\end{tabular}
\end{table}

\clearpage
\section{Experimental Design}

\begin{table}[htbp]
\centering
\caption{Overview of All Questionnaire Items and Measures}
\label{tab:all_measures}
\tiny % Smaller font to fit more content
\begin{tabular}{p{2.1cm}p{1.6cm}p{9.4cm}p{2.1cm}}
\toprule
\textbf{Construct} & \textbf{Response Format} & \textbf{Item Text} & \textbf{Reference}\\
\midrule
\multicolumn{4}{l}{\textbf{Pre-Task Measures}} \\
\midrule
General Self-Efficacy & 6-point Likert & 1. I am confident about my ability to do the task. \newline 2. I have mastered the skills necessary for the task. \newline 3. I am self-assured about my capabilities to perform the task. & \citet{spreitzer1995psychological}  \\
\addlinespace
General AI Efficacy & 6-point Likert & 1. I am confident in the AI's ability to do the task. \newline 2. The AI has the required skills for the task. \newline 3. I am convinced of the AI's capabilities to perform the task. & \citet{spreitzer1995psychological} \\
\addlinespace
Cognitive Reflection Test & Open-ended & 1. A bat and a ball cost \$1.10 in total. The bat costs \$1.00 more than the ball. How much does the ball cost? \newline 2. If it takes 5 machines 5 minutes to make 5 widgets, how long would it take 100 machines to make 100 widgets? \newline 3. In a lake, there is a patch of lily pads. Every day, the patch doubles in size. If it takes 48 days for the patch to \newline cover the entire lake, how long would it take for the patch to cover half of the lake? & \citet{frederick2005cognitive} \\
\addlinespace
Task Familiarity & 6-point Likert & How familiar are you with the task domain of loan approval decisions? & Self-developed\\
\midrule
\multicolumn{4}{l}{\textbf{Post-Task Measures}} \\
\midrule
General Self-Efficacy & 6-point Likert & Same three items as in the pre-task questionnaire. & \citet{spreitzer1995psychological} \\
\addlinespace
General AI Efficacy & 6-point Likert & Same three items as in the pre-task questionnaire. & \citet{spreitzer1995psychological} \\
\addlinespace
System 1 Processing & 6-point Likert & 1. I used my gut feelings to conduct the loan approval task. \newline 2. I knew my decision at first glance when looking at each case. \newline 3. I relied on my first impressions after screening the information and considering what I received from the AI. & \citet{novak2009fit} \\
\addlinespace
System 2 Processing & 6-point Likert & 1. I reasoned things out carefully during the loan approval task. \newline 2. I approached the loan approval task systematically. \newline 3. I arrived at my decisions by carefully assessing all the available information.  & \citet{novak2009fit} \\
\addlinespace
Learning from Errors & 6-point Likert & 1. During the study, I tried to recognize my errors to learn and improve from them. \newline 2. My errors during the study assisted me to improve my work on the loan approval task. \newline 3. Realizing my errors provided useful information for me to carry out my work on the loan approval task. & \citet{rybowiak1999error}\\
\addlinespace
Feedback Helpfulness & 6-point Likert & The feedback was helpful in making my subsequent decisions. & Self-developed \\
\addlinespace
Feedback Helpfulness & Open-ended & 1. How did the examples in the feedback influence your loan approval decisions? \newline 2. How did the information on the number of instances where you disagreed with the AI advice \newline in the feedback influence your loan approval decisions? \newline 3. Which elements of the feedback did you consider most and why? & \citet{he2023knowing} \\
\addlinespace
Demographics & Various & 1. Age  \newline 2. Gender \newline 3. Employment status \newline 4. Education level \newline 5. Country of residence & Self-developed\\
\bottomrule
\end{tabular}
\end{table}

\end{document}